\documentclass[
 reprint,
 preprintnumbers,
 amsmath,
 amssymb,
 aps,
 prstab,
 floatfix,
]{revtex4-1}

\usepackage{graphicx}					% Graphics
\usepackage{verbatim}					% Graphics
\usepackage{wasysym}					% Bold PI
\usepackage{mathtools}					% Curved arrow
\usepackage[makeroom]{cancel}				% Crossed math
\usepackage{dcolumn}		% Align table columns on decimal point
\usepackage{hyperref}		% Add hypertext capabilities
\usepackage{color}                    %  DO NOT REMOVE!!!

\newcommand{\pd}{\partial}				% Partial derivative
\newcommand{\dd}{\mathrm{d}}				% General derivative

\newcommand{\Nb}{N_{\text{b}}}		% N_b number of particles in bunch
\newcommand{\Qb}{ Q_{\beta}}		% Q_b betatron tune
\newcommand{\Qs}{ Q_{\text{s}}}		% Q_s syncrotron tune
\newcommand{\Qef}{Q_{\text{eff}}}	% Q_eff tune shift
\newcommand{\Qmx}{Q_{\text{max}}}	% Q_max SC tune shift
\newcommand{\Qw}{ Q_{\text{W}}}		% Q_w wake-driven coherent tune shift

\newcommand{\Qsc}{\Delta Q_\mathrm{sc}}
\newcommand{\Qx }{\Delta Q_k          }
\newcommand{\qx }{\Delta q_k          }
\newcommand{\tb }{    \tau_\mathrm{b }}
\newcommand{\vv  }{         \mathrm{v }}
\newcommand{\kt  }{\kappa_\mathrm{TMCI}}

\newcommand{\Yk}{\mathcal{Y}_k}				% modes Y_k

\newcommand{\hG}{\widehat{\mathrm{G}}}			% matrix G
\newcommand{\hW}{\widehat{\mathrm{W}}}			% matrix W
\newcommand{\hD}{\widehat{\mathrm{D}}}			% matrix D
\newcommand{\mG}{\widehat{\mathrm{G}}_{lm}}		% matix element П_lm
\newcommand{\mW}{\widehat{\mathrm{W}}_{lm}}		% matix element Ц_lm
\newcommand{\mD}{\widehat{\mathrm{D}}_{lm}}		% matix element D_lm
\begin{document}

%===============================================================================
\title{TMCI and Space Charge}
\author{T.~Zolkin}
\email{zolkin@fnal.gov}
\affiliation{Fermilab, PO Box 500, Batavia, IL 60510-5011}
%\thanks{Operated by Fermi Research Alliance, LLC under Contract
%No.~De-AC02-07CH11359 with the U.S. Department of Energy.}
\author{A.~Burov}
\affiliation{Fermilab, PO Box 500, Batavia, IL 60510-5011}
\date{\today}
%===============================================================================

%===============================================================================
\begin{abstract}
Transverse mode-coupling instability (TMCI) is known to limit bunch intensity.
Since space charge (SC) changes the spectra of the collective modes, it affects
the TMCI threshold as well.
In agreement with results of M.~Blaskiewicz and V.~Balbekov, we found that, when
the wake is negative or {\it essentially negative}, the instability threshold
increases as fast as the space charge tune shift when the latter is large 
enough.
In contrast, for oscillating wakes, when the oscillations are sufficiently
pronounced, the threshold dependence on SC is non-monotonic;
at sufficiently high SC tune shift the threshold goes inversely proportional to 
that.
\end{abstract}
%===============================================================================

\pacs{00.00.Aa ,
      00.00.Aa ,
      00.00.Aa ,
      00.00.Aa }% PACS, the Physics and Astronomy Classification Scheme.
\keywords{Suggested keywords}% Use showkeys class option if keyword
                             % display desired

%===============================================================================
\maketitle
%\tableofcontents
%===============================================================================

%==============================================================================%
%==============================================================================%
%==============================================================================%
\section{Introduction}

%------------------------------------------------------------------------------%
A problem of coherent beam stability is known to be extremely hard when the beam
space charge (SC) has to be taken into account, which is typical for low- and 
medium-energy hadron rings.
Up to now, the only universally accurate method available is a multi-particle 
tracking;
since up to millions of macroparticles per bunch are needed for reliable
results, this method is very expensive in terms of CPU time.
This is why approximate analytical models are valuable: although each of them
has its limitations, they help to build general understanding, providing
important results within their areas of validity.
Since these areas, as well as the models accuracy,  are not always clear, 
results of available analytical models should be compared whenever possible.

%------------------------------------------------------------------------------%
Transverse mode coupling instability (TMCI), also known as strong head-tail
instability, is one of main intensity limitations of bunched beams in circular
machines \cite{chao1993physics}.
Specifics and even existence of this instability for beams with strong space 
charge, when the SC tune shift exceeds the synchrotron tune, remained not quite 
clear for rather long time since TMCI without SC was principally understood and 
described.
A publication with a significant breakthrough in this direction, suggested by 
M.~Blaskiewicz, appeared about twenty yeas ago \cite{blaskiewicz1998fast}.
A model of an airbag bunch within a square potential well (addressed here as ABS
model) has been presented there and analyzed with an exponential wake and, in 
principle, arbitrary ratio of the SC tune shift to the synchrotron tune, or 
{\it the space charge parameter}.
A method to extend this model to an arbitrary sum of exponential/oscillating 
wakes was also described.
It was shown, that for wide range of the wake decay rate and the SC parameter, 
the instability threshold grows with the latter.
A qualitative explanation why SC may work this way was suggested next year 
\cite{ng1999stability}.
In the year of 2009, theory of head-tail instabilities with strong space charge 
(SSC) was presented in Refs.~\cite{burov2009head,balbekov2009transverse}.
One of the authors of this paper speculated then that dependence of the TMCI 
threshold on the SC parameter might be non-monotonic, the threshold might
start to decrease after a certain value of the SC tune shift.
It was demonstrated that for cosine wake this statement is true, but whether it 
is true for sign-constant wakes remained unclear.
About a year ago, V.~Balbekov confirmed his agreement \cite{balbekov2016tmci} 
with this hypothesis for the constant wake and {\it boxcar model} (referred as 
HP$_0$ or simply HP distribution in this paper).
Soon after that, however, he withdrew this result, claiming that for both ABS 
and boxcar models the threshold grows monotonically with the space charge 
parameter, without any sign to change this trend at higher SC 
\cite{balbekov2016dependence,balbekov2016tmci,balbekov2017transverse,
balbekov2017threshold}.
It has been also shown \cite{balbekov2017transverse} that for the cosine 
wake with the phase advance not exceeding $\pi$, the TMCI threshold 
monotonically increases with the SC parameter, qualitatively similar to the 
negative wakes.
%We confirmed that for both SSC models there is no TMCI when phase advance  
%$<\pi$, however, when phase advance is larger than $\pi$, we show that there is
%a non-vanishing TMCI.
%Using examples of the cosine wake with phase advance in the range 
%$2\,\pi$--$6\,\pi$ we demonstrate that the behavior of TMCI threshold in the
%ABS model might be not monotonic as a function of SC, and for sufficiently 
%large SC parameter the threshold goes to zero.

%------------------------------------------------------------------------------%
In this paper, we are trying to get a better confidence and wider vision of the 
TMCI threshold versus the SC parameter for various wakes, potential wells and 
bunch longitudinal distributions.
Our paper, similar to \cite{balbekov2017transverse,balbekov2017threshold}, 
is a summary of numerical results for several models. 

%------------------------------------------------------------------------------%
We use two analytical approaches, the ABS of M.~Blaskiewicz and the strong 
space 
charge theory of Ref.~\cite{burov2009head}, SSC.
While the ABS model can be applied for any SC strength, the SSC theory is 
applicable only when the SC tune shift sufficiently exceeds both the 
synchrotron 
tune and the wake's one, or the coherent tune shift. However, the SSC theory 
has 
an advantage in its applicability for any potential wells and longitudinal 
distributions, so the two approaches complement each other.
Within the SSC approach, we, as V.~Balbekov, examine the square potential well 
with arbitrary longitudinal distribution, and the boxcar distribution for the 
parabolic potential well.
The former of them we call as SSCSW (strong space charge, square well), and the 
latter we prefer to address as the generalized Hofmann-Pedersen distribution of 
0 order, HP$_0$, or  SSCHP model.
The reason for this name change is that for all the models under consideration, 
the bunch line density is boxcar, or constant, so the term {\it boxcar} for a 
specific model may 
lead to a confusion.       

%------------------------------------------------------------------------------%
Recently, one more model for a bunch with SC and a constant wake has been
suggested, namely, the two-particle one \cite{chin2016two}.
We think that its main advantage, simplicity, is somewhat excessive for
the specific problem under study, so we leave its possible application outside
the framework of this paper. 

%------------------------------------------------------------------------------%
For the negative exponential wakes, our conclusion agrees with M.~Blaskiewicz
\cite{blaskiewicz1998fast} and the last results of V.~Balbekov
\cite{balbekov2017transverse,balbekov2017threshold}: SC always elevates the 
TMCI threshold; when the SC parameter is large, the threshold grows linearly 
with that.  

%------------------------------------------------------------------------------%
For the oscillating wakes and strong space charge, the situation is opposite, 
the SC makes the beam less stable: the threshold is found to be 
inversely proportional to the SC parameter when the wake phase advance is large 
enough, $\omega\,\tb>\pi$. For the same case with small or moderate SC 
parameter, though, the thresholds typically show certain growth with SC, thus 
confirming the hypothesis of Ref.~\cite{burov2009head} about non-monotonic 
dependence of the TMCI threshold on the SC parameter for oscillating wakes.

When the wake oscillations are not pronounced (either the phase 
advance is insufficient, $\omega\,\tb < \pi$, or the wake oscillations are 
overshadowed by their exponential decay) the TMCI vanishes at the SSC limit, as 
it can be expected.
For the SSCHP model and the sine wake, the modes behave in an unexpected way: 
they cross or approach and then divert away from each other, always without 
coupling, as if the instability is somehow prohibited for this specific case.

%------------------------------------------------------------------------------%
%It was demonstrated that fast head-tail instability appears when the 
%betatron tune shift becomes comparable to the synchrotron tune.
%------------------------------------------------------------------------------%

%==============================================================================%
%==============================================================================%
\subsection{Article structure}

%------------------------------------------------------------------------------%
The paper is structured as follows.
Sec.~\ref{sec:Models} summaries the SSC and ABS main formulas for a single 
bunch at zero chromaticity, for the reader's convenience.
Modes for the no-wake case, or the {\it SSC harmonics}, are presented for two
cases, for a bunch in a square well and for the $\mathrm{HP}_0$ distributions.
Subsection~\ref{sec:AB} describes the airbag model.
Sections~\ref{sec:NegWakes} and \ref{sec:OscWakes} are dedicated to negative
(delta-functional, constant, exponential and resistive wall) and oscillating
(sine, cosine and broadband impedance) wake functions respectively.
The results are summarized in Sec.~\ref{sec:RnD}.
Appendix~\ref{secAP:Wlm} contains the wake matrix elements for the SSCHP model.
Appendix~\ref{secAP:AB} contains details on the exponential and trigonometric 
wakes for the ABS model.

%==============================================================================%
%------------------------------------------------------------------------------%
%------------------------------------------------------------------------------%
%------------------------------------------------------------------------------%
%==============================================================================%
\section{\label{sec:Models}Analytical models}

%==============================================================================%
%------------------------------------------------------------------------------%
%==============================================================================%
\subsection{\label{sec:SSCTheory}Strong space charge theory}

%------------------------------------------------------------------------------%
In this subsection, we remind main formulas of the SSC theory 
\cite{burov2009head}.
Consider a single bunch with a longitudinal distribution function $f(\tau,\vv)$,
where $\tau$ is the position along the bunch, and $\vv$ is
the particle longitudinal velocity, $\vv=d\tau /d \theta$; both the position 
$\tau$ and time $\theta$ are measured in radians.
For zero wake, transverse modes satisfy an ordinary differential equation and 
zero-derivative boundary conditions, which constitutes a standard 
Sturm-Liouville (S-L) problem,
\begin{equation}
\left\{\begin{array}{l}\displaystyle
	\frac{1}{\Qef(\tau)}\frac{\dd}{\dd\tau}
	\left(
		u^2(\tau) \frac{\dd\,Y(\tau)}{\dd \tau}
	\right)
	 + \nu\,Y(\tau)
	= 0,					\\[0.5cm]
	\displaystyle
	\left.\frac{\dd}{\dd\tau}\,Y(\tau) \right|_{\tau = \pm\infty} = 0.
\end{array}\right.
\end{equation}
Solutions of this problem constitute an orthogonal basis; being normalized, they 
are referred as 
the {\it SSC harmonics} $[\nu_k,Y_k(\tau)]$:
\begin{equation}
	\int_\text{SB}\rho(\tau)\,Y_l(\tau)\,Y_m(\tau)\,\dd\tau = \delta_{lm},
\end{equation}
with $\rho(\tau)$ being the normalized line density
\begin{equation}
\label{math:NormCond}
	\rho(\tau) = \int_\text{SB} f(\tau,\vv) \,\dd\vv:\qquad
	\int_\text{SB} \rho(\tau)\,\dd\tau = 1;
\end{equation}
the integrals are taken along the bunch length (SB stays for a single bunch).
The temperature function $u^2(\tau)$ is the local average of the longitudinal
velocity squared,
\begin{equation}
	u^2(\tau) = \int f(\tau,\vv)\,\vv^2\,\dd \vv
	\,\bigg/\,\rho(\tau);
\end{equation}
$\Qef$ is the {\it effective space charge tune shift} at the given position 
along the 
bunch,
\begin{equation}
	\Qef(\tau) = \Qef(0)\,\frac{\rho(\tau)}{\rho(0)},
\end{equation}
where 'effective' means the transverse average at the given position $\tau$ 
along the bunch.
If the transverse distribution is Gaussian, the effective space charge tune 
shift is $\approx 0.52$ 
of its value at the beam axis. 

%------------------------------------------------------------------------------%
Wake function $W(\tau)$ modifies the collective 
dynamics as follows:
\begin{equation}
	\frac{1}{\Qef(\tau)}\frac{\dd}{\dd\tau}
	\left(
		u^2(\tau) \frac{\dd\,\mathcal{Y}(\tau)}{\dd \tau}
	\right)
	+ \Delta q\,\mathcal{Y}(\tau)
	= \kappa\,\hW\,\mathcal{Y}(\tau), 
\end{equation}
% [ \varkappa( \hW + \hD ) - i\,g\,e^{i\,\psi}\,\hG ] \mathcal{Y}(\tau)
with
\begin{equation}
	\hW\,Y(\tau) = \int_{\tau}^\infty
	W(\tau-\sigma)\,\rho(\sigma)\,Y(\sigma)\,
	\dd\sigma,
%\hD\,Y(\tau) = Y(\tau)
%	\int_{\tau}^\infty
%		D(\tau-\sigma)\,\rho(\sigma)\,
%	\dd\,\sigma
\end{equation}
and
% $\zeta = -\xi/\eta$ is the negated ratio of conventional chromaticity, $\xi$,
% and a slippage factor, $\eta = \gamma_\text{t}^{-2}-\gamma^{-2}$.
\begin{equation}
	\kappa = \Nb\,\frac{r_0 R_0}{4\,\pi\,\gamma\,\beta^2\,\Qb}.
\end{equation}
Here $\Nb$ is the number of particles per bunch,
$r_0$ is the particle classical radius,
$R_0=C_0/(2 \pi)$ is the average accelerator ring radius,
$\Qb$ is the bare betatron tune, 
$\gamma$ is the Lorentz factor
and
$\beta$ is the ratio of particle velocity to the speed of light.

%------------------------------------------------------------------------------%
\begin{figure*}[t!]
\includegraphics[width=\linewidth]{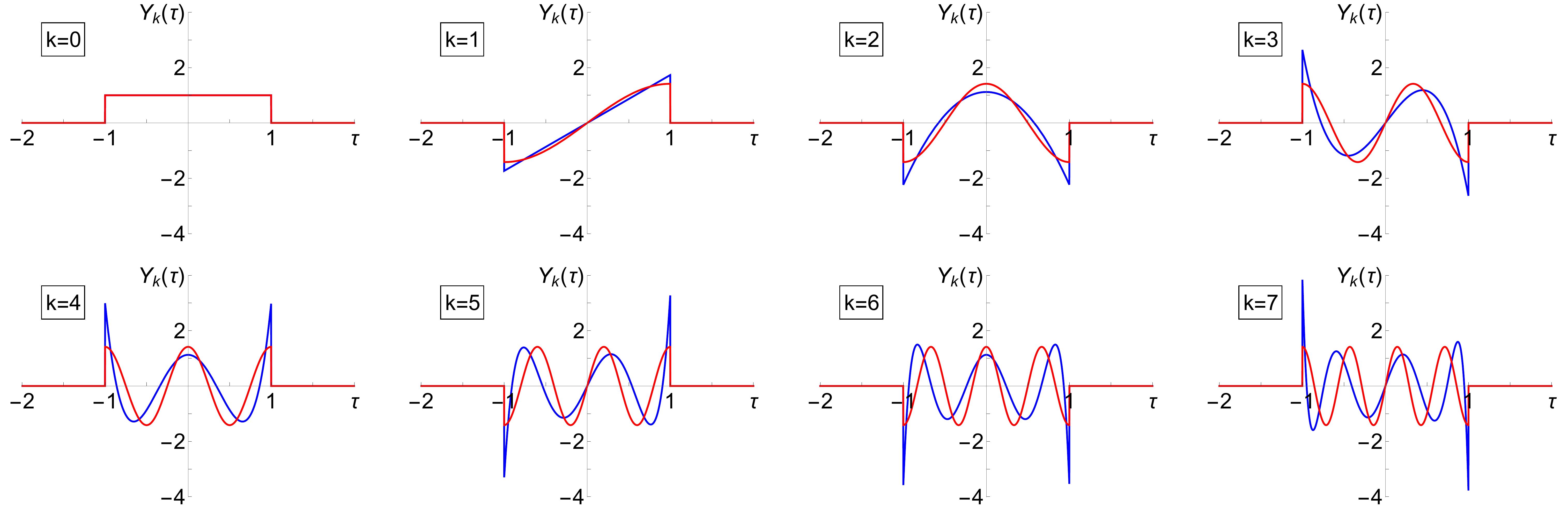}
\caption{\label{fig:SSCHarm}
	First eight SSC harmonics $Y_k(\tau)$ for a bunch in a square potential 
	well (red) and $\text{HP}_0$ distribution (blue).
	The harmonics are presented in dimensionless units.
	Note that for both distributions $\rho(\tau) = 1/2$.	}
\end{figure*}
%------------------------------------------------------------------------------%

%------------------------------------------------------------------------------%
Expansion over the SSC harmonics
\begin{equation}
	\Yk(\tau) = \sum_{i=0}^{\infty} \mathbf{C}^{(k)}_i Y_i(\tau)
\end{equation}
leads to the eigenvalue problem
$\mathbf{M}\cdot\mathbf{C}^{(k)} = \qx \mathbf{C}^{(k)}$
where
\begin{equation}
	\mathbf{M}_{lm} =
	\nu_l\,\delta_{lm} +
	      \kappa\,\mW
\end{equation}
with matrix elements of the wake operator being
\begin{equation}
	\mW = \int_{-\infty}^\infty \int_{\tau}^\infty
		W(\tau-\sigma)\,
		\rho(\tau) \, \rho(\sigma) \, Y_l(\tau) \, Y_m(\sigma)
		\,\dd\,\sigma\,\dd\,\tau.
%\mD &=& \int_{-\infty}^\infty \int_{\tau}^\infty
%	\dd\,\sigma\,\dd\,\tau\,D(\tau-\sigma)\,
%		\rho(\tau) \, \rho(\sigma) \, Y_l(\tau) \, Y_m(\tau)\,,	\\
%\mG &=& \mathrm{K}_l(\zeta)\,\mathrm{P}^*_m(\zeta)\,, 
%\quad\text{where}							\\
%\mathrm{P}_k\,\, &=& \int_{-\infty}^\infty\,\dd\,\tau\,
%	P(\tau) \, \rho(\tau) \, Y_k(\tau) \, e^{i\,\zeta\,\tau}\,,	\\
%\mathrm{K}_k\,\, &=& \int_{-\infty}^\infty\,\dd\,\tau\,
%	K(\tau) \, \rho(\tau) \, Y_k(\tau) \, e^{i\,\zeta\,\tau}\,.
\end{equation}
Below, real and imaginary parts of the eigenvalues are denoted as
\begin{equation}
	\qx = \Delta_k + i\,\Gamma_k.
\end{equation}

%------------------------------------------------------------------------------%
As it was demonstrated by V.~Balbekov \cite{balbekov2017threshold}, the
expansion over the SSC harmonics may have a poor convergence, so that the
numerically obtained instability threshold is very sensitive to the accuracy of 
the matrix elements computation and the number of
harmonics taken into account.
In order to distinguish the illusory, pure numerical, threshold from the real
one, two things are required: a large number of the basis harmonics and high 
accuracy of the computed matrix elements.
To provide that, we considered two different distributions with constant line 
density, where the matrix elements can be computed analytically, thus minimizing
numerical errors up to machine precision and justifying to include large number 
of modes into analysis.

%==============================================================================%
%==============================================================================%
\subsubsection{\label{sec:SSCSW}Square potential well}

%------------------------------------------------------------------------------%
Our first SSC model relates to a bunch in the square potential well 
(SSCSW, stays for Strong Space Charge, Square Well).
In this case the distribution function is factorized
\begin{equation}
	f_\text{sw}(\tau,\vv) =
		\frac{1}{2\,a}\text{H}\left[ 1-\frac{\tau^2}{a^2}\right]
		\mathrm{V}(\vv).
\end{equation}
Below we omit the Heaviside function, assuming that all bunch related functions 
are defined on its length, $|\tau| \leq a$.
The average square of velocity is position-independent, 
\begin{equation}
	u_\text{sw}^2(\tau) \equiv b^2.
\end{equation}

%------------------------------------------------------------------------------%
Then, the equation for the SSC harmonics can be written as
\begin{equation}
	0 = \nu\,Y_k^\text{sw}(\tau) +
		\frac{b^2}{\Qef(0)}\frac{\dd}{\dd\tau}\left[
		\frac{\dd}{\dd\tau}\,Y_k^\text{sw}(\tau)
	\right].
\end{equation}
With $\tau$ measured in units of $a$ and $\nu$ in units of 
$\pi^2b^2/[4\,a^2\Qef(0)]$, it reduces to
\begin{equation}
\left\{
	\begin{array}{l}\displaystyle
		Y''(\tau) + \frac{\pi^2}{4}\,\nu\,Y(\tau) = 0,	\\[0.25cm]
		Y'(\pm 1) = 0.
	\end{array}
\right.
\end{equation}
This S-L problem yields the eigenvalues as integers squared,
\begin{equation}
	\nu_k^\text{sw} = \left\{k^2\,|\,k\in\mathbb{Z} \right\} =
	\left\{ 0,1,4,9,25,36,\ldots \right\}.
\end{equation}
The set of eigenfunctions consists of the constant 0-th mode $Y^\text{sw}_0 = 1$
and a sequence of cosine and sine functions (see Fig.~\ref{fig:SSCHarm})
\begin{equation}
Y_k^\text{sw}(\tau) = \left\{
\begin{array}{lll}
\displaystyle \sqrt{2-\delta_{0k}}\,\cos\frac{\pi\,k\,\tau}{2},	&
\qquad & k\text{ is even},					\\[0.25cm]
\displaystyle \sqrt{2}\,\sin\frac{\pi\,k\,\tau}{2},		&
\qquad & k\text{ is odd}.
\end{array}
\right.
\end{equation}
Note that the functions are normalized according to the general SSC harmonics 
orthonormalization rule~(\ref{math:NormCond}) with $\rho(\tau)=\frac{1}{2}$.

%==============================================================================%
%==============================================================================%
\subsubsection{\label{sec:SSCHP}Hofmann-Pedersen distribution}

%------------------------------------------------------------------------------%
Another SSC model, where matrix elements for some wakes can be analytically 
calculated, is a bunch with a constant line density inside a parabolic potential
well.
Following M.~Blaskiewicz, this distribution was previously referred as a
{\it boxcar}, but since all our models have a boxcar, or constant, line density, 
in order to avoid confusions,
we refer to it as the $\text{HP}_0$ model, or zeroth Hofmann-Pedersen 
distribution.
Its phase space density is
\begin{equation}
f_\text{hp}^{(0)}(\tau,\vv) = 
	\frac{\text{H}\left[ 1-\frac{\tau^2}{a^2}-\frac{v^2}{b^2} \right]}
	{2\,\pi\,a\,b}
	\left(1-\frac{\tau^2}{a^2}-\frac{v^2}{b^2}\right)^{-1/2}.
\end{equation}
Normalized line density and temperature function are computed as
\begin{equation}
	\rho(\tau) = \frac{1}{2},\qquad\qquad
	u_\text{hp}^2(\tau) = \frac{b^2}{2}\left( 1-\frac{\tau^2}{a^2}\right).
\end{equation}
This notation assumes a generalized Hofmann-Pedersen distribution of $n$-th 
order,
\begin{equation}
	f_\text{hp}^{(n)} \propto (1-\tau^2/a^2 - v^2/b^2)^{n-1/2},
\end{equation}
so the conventional Hofmann-Pedersen distribution~\cite{hofmann1979bunches}, 
with its parabolic line density, can be addressed as $f_\text{hp}^{(1)}$.
In this paper, only zero-order case $\text{HP}_0$ is considered, so the 
subscript $_0$ can be safely omitted.

%------------------------------------------------------------------------------%
The equation for SSC harmonics
\[
    \nu\,Y_k^\text{hp}(\tau) +
    \frac{b^2}{2\,\Qef(0)}\frac{\dd}{\dd\tau}\left[
    \left( 1-\frac{\tau^2}{a^2}\right)\frac{\dd}{\dd\tau}\,Y_k^\text{hp}(\tau)
    \right] = 0
\]
can be conveniently written in the normalized variables, when $\tau$ is 
measured 
in units of $a$ and $\nu$ in units of $b^2/[a^2\Qef(0)]$, 
\begin{equation}
	\label{math:HPHarm}
	\frac{\dd}{\dd\tau}\left[
	\left( 1-\tau^2 \right)\frac{\dd}{\dd\tau}\,Y_k^\text{hp}(\tau)
	\right] + 2\,\nu\,Y_k^\text{hp}(\tau) = 0.
\end{equation}
Note that $\nu$ is measured in different units in SSCSW and SSCHP;
this choice was made in order to have first eigenvalues
$\nu_1 = 1$ for the both models.

%------------------------------------------------------------------------------%
For this distribution, the S-L problem doesn't have 
solutions, which relates to the singularities in
Eq.~(\ref{math:HPHarm}) at $\tau = \pm 1$.
One way to overcome this problem is to add a small regularization term 
$\epsilon>0$,
\begin{equation}
	Y''(\tau)
	- \frac{2\,\tau}{1+\epsilon-\tau^2}\,Y'(\tau)
	+ \frac{2\,\nu }{1+\epsilon-\tau^2}\,Y (\tau)
	= 0,
\end{equation}
thus removing the singularity within the same bunch length $[-1;1]$.
With this regularization, the S-L solutions do exist, and, as
$\epsilon \rightarrow 0$, the eigenvalues converge to the {\it triangular 
numbers}
\begin{equation}
	\nu_k^\text{hp} = \left\{k\,(k+1)/2\,|\,k\in\mathbb{Z} \right\} =
	\left\{ 0,1,3,6,10,15,\ldots \right\}.
\end{equation}
The eigenfunctions converge to ones proportional to the Legendre 
polynomials $P_k(\tau)$, 
\begin{equation}
Y_k^\text{hp} = 
	(-1)^{\left\lfloor{k/2}\right\rfloor}\sqrt{2\,k+1}\,P_k(\tau)
\end{equation}
where the floor function $\left\lfloor n \right\rfloor$ is an integer part of 
$n$, and $(-1)^{\left\lfloor k/2 \right\rfloor}$ factor guarantees that each 
even (odd) harmonic has cosine-like (sine-like) properties at the origin, see 
Eq.~(\ref{math:SC-like}). These SSC harmonics are presented in 
Fig.~\ref{fig:SSCHarm}.

%------------------------------------------------------------------------------%
Another way to find eigenvalues is to look for an even and odd functions with
\begin{equation}
\label{math:SC-like}
	\left\{
	\begin{array}{l}
		Y\phantom{'}(0)	= 1,		\\
		Y'(0)		= 0,
	\end{array}
	\right.
	\qquad\text{or}\qquad
	\left\{
	\begin{array}{l}
		Y\phantom{'}(0)	= 0,		\\
		Y'(0)		= 1,
	\end{array}
	\right.
\end{equation}
respectively.
When the small regularization term $\epsilon > 0$, the derivatives $Y'(\pm 1)$ 
vanish for the same triangular 
values, $\nu_k^\text{hp}$.
With $\epsilon=0$, only the triangular eigenvalues $\nu=k(k+1)/2$ yield the 
solutions, remaining finite at the bunch edges $\tau = \pm 1$.
Analytical expressions for the matrix elements $\mW$ are provided in
Appendix~\ref{secAP:Wlm}.

%==============================================================================%
%------------------------------------------------------------------------------%
%==============================================================================%
\subsection{\label{sec:AB}Airbag in a square well}

%------------------------------------------------------------------------------%

In addition to SSC cases, we consider the airbag longitudinal distribution
inside a square well, abbreviated as ABS (AirBag Square well) model,
\begin{equation}
	f_\mathrm{ABS}(\tau,\vv) \propto
	\left[ \delta(\vv-\vv_0) + \delta(\vv+\vv_0) \right]
	\text{H}\left[ 1-4\,\frac{\tau^2}{\tb^2}\right],
\end{equation}
suggested by M.~Blaskiewicz \cite{blaskiewicz1998fast}.
For a wide class of wake functions, the bunch spectrum can be found for
arbitrary space charge tune shift, without expansion over an infinite set of 
basis function, thus avoiding the related convergence problem.

%------------------------------------------------------------------------------%
It is convenient to take for this model the following convention
$\tau\in[-\tb/2;\tb/2]$ for the position along the bunch, where all particles 
move with the same absolute value of the velocity,  
$\dd\,\tau/\dd\theta = \pm \vv_0$.
Transverse offsets in the two fluxes of particles $X_\pm(\theta,\tau)$ can be 
looked for as
\begin{equation}
	X_\pm(\theta,\tau) = e^{-i\,(\Qb+\Qx)\theta} x_\pm(\tau)
\end{equation}
where$\Qb$ is the bare betatron tune and $\Qx$ is the tune shift to be found. 
Then, equations for the amplitudes along the bunch $x_\pm(\tau)$ are given by
\begin{eqnarray}
\label{math:xp}
	\frac{\dd\,x_+}{\dd \tau} & = & \frac{i}{\vv_0} \left[
		\left( \frac{\Qsc}{2}+\Qx \right) x_+ - \frac{\Qsc}{2}\,x_- - F
	\right],\qquad						\\
\label{math:xm}
	\frac{\dd\,x_-}{\dd \tau} & = & \frac{i}{\vv_0} \left[
		\frac{\Qsc}{2}\,x_+ - \left( \frac{\Qsc}{2}+\Qx \right) x_- + F
	\right],\qquad
\end{eqnarray}
with the boundary conditions
\begin{equation}
\begin{array}{c}
	x_+( \tb/2) = x_-( \tb/2),	\\[0.2cm]
	x_+(-\tb/2) = x_-(-\tb/2).
\end{array}
\end{equation}

%------------------------------------------------------------------------------%
The force is defined by a wake function
\begin{equation}
	F(\tau) = \kappa\,\int_\tau^{\tb/2}
		W(\tau-\sigma)\,\overline{x}(\sigma)
	\,\dd\sigma
\end{equation}
and satisfies an integro-differential equation:
\begin{equation}
	\label{math:F(t)}
	\frac{\dd\,F(\tau)}{\dd \tau} =-\kappa\,W(0)\,\overline{x}(\tau) +
		\kappa\,\int_\tau^{\tb/2} \frac{\pd}{\pd \tau} W(\tau-\sigma) 
	\overline{x}(\sigma)
		\,\dd \sigma
\end{equation}
with $\overline{x} = (x_+ + x_-)/2$.
For a wake function in the form
\begin{equation}
\label{math:W(tau)}
	W(\tau) = -W_0\,\sum_{k=1}^{n} C_k\,e^{\alpha_k\,\tau}
\end{equation}
the force is
\begin{equation}
	F = \sum_{k=1}^n F_k =-\kappa\,W_0  \sum_{k=1}^n  C_k \int_\tau^{\tb/2} 
	e^{\alpha_k(\tau-\sigma)}\,\overline{x}(\sigma)\,\dd \sigma,
\end{equation}
and for every $k=1,\ldots,n$,
\begin{equation}
\label{math:dFk}
	\frac{\dd\,F_k(\tau)}{\dd \tau} =
	\frac{\kappa\,W_0}{2}\,C_k(x_+ + x_-) + \alpha_k\,F_k,
\end{equation}
with $F_k(\tb/2) = 0$.

%------------------------------------------------------------------------------%
Measuring $\tau$ in units of $\tb$ and defining
\begin{equation}
	f(\tau) = F(\tau)/\Qs
\end{equation}
where $\Qs = \pi\,\vv_0/\tb$ is the synchrotron tune,
Eqs.~(\ref{math:xp},\ref{math:xm}) and (\ref{math:dFk}) can be presented as a 
set of 
ordinary linear homogeneous differential equations,
\begin{equation}
	\frac{\dd\,\mathrm{U}}{\dd\tau} = \mathrm{M}\,\mathrm{U},
	\label{dUdtau}
\end{equation}
with boundary conditions
\begin{eqnarray}
	f_k( 1/2) &=& 0,			\\
	x_+( 1/2) &=& x_-( 1/2),		\\
	x_+(-1/2) &=& x_-(-1/2),
\end{eqnarray}
where $\mathrm{U} = (x_+,x_-,f_1,f_2,\ldots)$, and the matrix $\mathrm{M}$ is 
combined from the coefficients.
% The general solution, except for special cases, is
% \begin{equation}
%	\mathrm{U} = \sum_{k=1}^n \mathcal{C}_k \mathrm{U}_k e^{\lambda_k\tau}.
% \end{equation}
% M.~Blaskiewicz suggested to choose a starting value of $\Qx$ and iterate until
% it corresponds to a solution that satisfies the boundary conditions.
To solve this set of equations, we suggest an algorithm different from one 
applied 
by M.~Blaskiewicz.

%------------------------------------------------------------------------------%
The differential Eq.~(\ref{dUdtau}), first, can be transformed into a 
difference 
one,
\begin{equation}
	\mathrm{U}_{n+1} - \mathrm{U}_n = \Delta\tau\,\mathrm{M}\,\mathrm{U}_n.
\end{equation}
Choosing $n=2^m$ and $\Delta\tau = 1/2^m$, for
$\mathrm{U}_n \equiv \mathrm{U}(1/2) = (1,1,0,0,\ldots)$ one has
\begin{equation}
	\mathrm{U}_0 \equiv \mathrm{U}(-1/2) = \left(
		\frac{1}{2^m}\,\mathrm{M} + \mathrm{I} 
	\right)^{-2^m}\,\mathrm{U}_n.
\end{equation}
Those values of $\Qx$, with which the boundary condition at the tail of the 
bunch is satisfied, i.e. 
\begin{equation}
	x_+^{(0)} - x_-^{(0)}=0 ,
\end{equation}
constitute the bunch spectrum; they can be found by a proper scan of the 
complex 
plane of $\Qx$.
However, for the instability thresholds, it is sufficient to scan only real 
$\Qx$: the threshold can be detected as reduction of the number of these real 
modes by two.

%------------------------------------------------------------------------------%
This algorithm proves to be extremely time-efficient; the solution always
converges with the number of intervals $1/\Delta \tau$, while the CPU time 
grows 
only logarithmically, $\propto \log_2 (1/\Delta \tau) =m$.

\begin{figure}[b!]
\includegraphics[width=\linewidth]{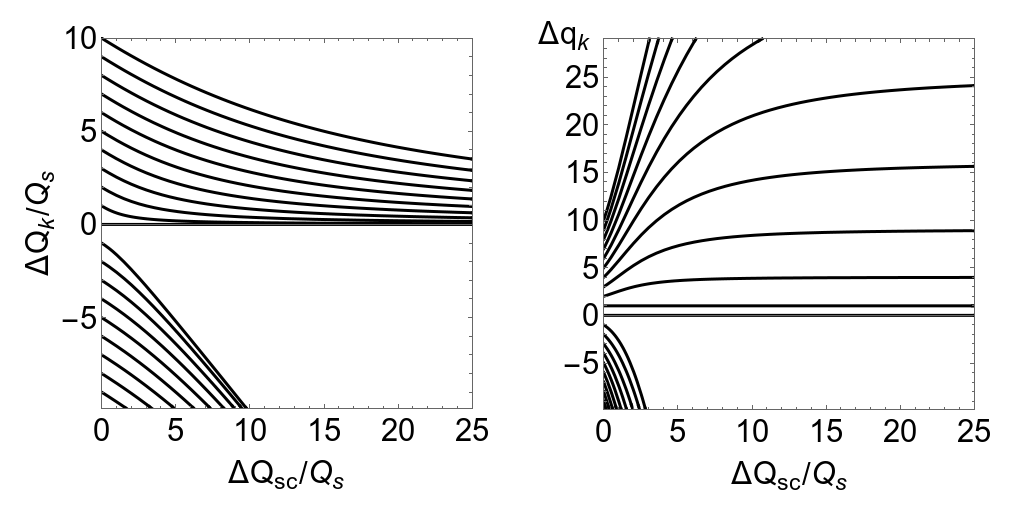}
\caption{\label{fig:ABNoWake}
	Eigenvalues (left) and normalized eigenvalues (right) for the ABS model 
as 
	functions of the SC parameter for no-wake case, $k=-9,\ldots,10$.
	}
\end{figure}

For the no-wake case, the ABS spectrum is calculated as
\begin{equation}
	\frac{\Qx}{\Qs} = -\frac{\Qsc}{2\,\Qs} \pm \
	\sqrt{\left(\frac{\Qsc}{2\,\Qs}\right)^2 + k^2},
\end{equation}
for $k = \pm 1,\pm 2,\ldots$ and additional zeroth mode $\Delta Q_0 = 0$ which 
is not affected by the SC.
When the SC becomes strong,
\begin{equation}
\label{math:SSC}
	\Qsc \gg 2\,k\,\Qs 
\end{equation}
the equidistant spectrum of integers (for $\Qsc=0$) separates on positive part,
quadratic with $k$: $\Qx \propto k^2$, and negative part, with $\Qx \approx 
-\Qsc$ (see Fig.~\ref{fig:ABNoWake}).
In order to make a comparison with the SSC case, we use normalized tunes
\begin{equation}
	\qx = \Qx / \Delta Q_1,
\end{equation}
so that the distance between the first and zeroth modes is equal to one for any 
value
of the SC parameter, $\Qef(0)/\Qs$, when there is no wake.
When condition~(\ref{math:SSC}) is satisfied, ABS positive spectrum coincides 
with the SSCSW one:
\begin{equation}
	\Qx = k^2 \Qs^2/\Qsc , \qquad k=0,1,2,\ldots. ;
\end{equation}
for the SSCSW model, all the details of the longitudinal phase space density 
may 
play a role only by means of the average synchrotron frequency $\Qs$.

%------------------------------------------------------------------------------%
%------------------------------------------------------------------------------%

\newpage

%------------------------------------------------------------------------------%
\begin{figure*}[h!]
\includegraphics[width=\linewidth]{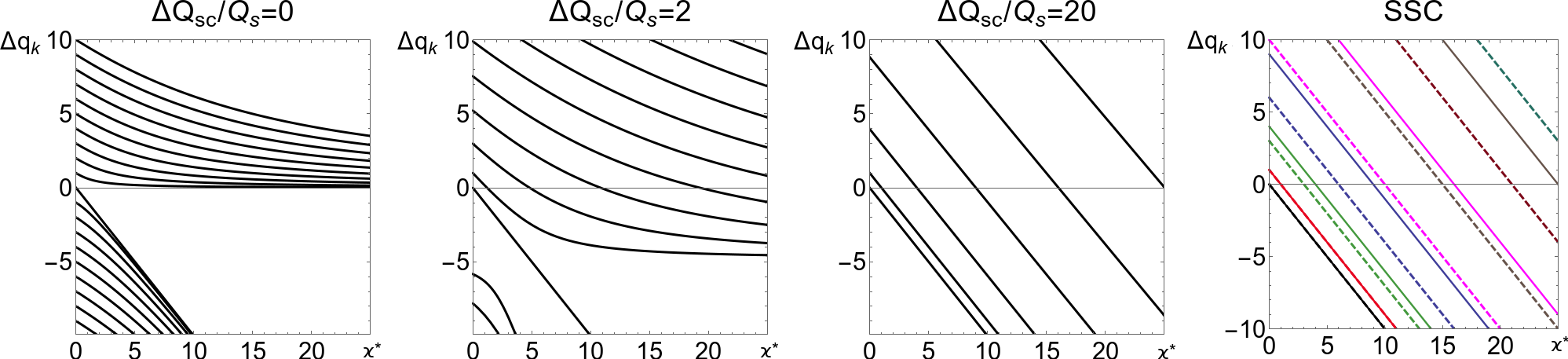}
\caption{\label{fig:DeltaWake}
	Normalized tune shifts for the ABS model and delta wake for different 
	values of the SC parameter (first three plots).
	The last figure shows same values for the SSCSW (solid lines) and 
	SSCHP (dashed lines) models.
	}
\end{figure*}
%------------------------------------------------------------------------------%

%------------------------------------------------------------------------------%
\begin{figure*}[h!]
\includegraphics[width=\linewidth]{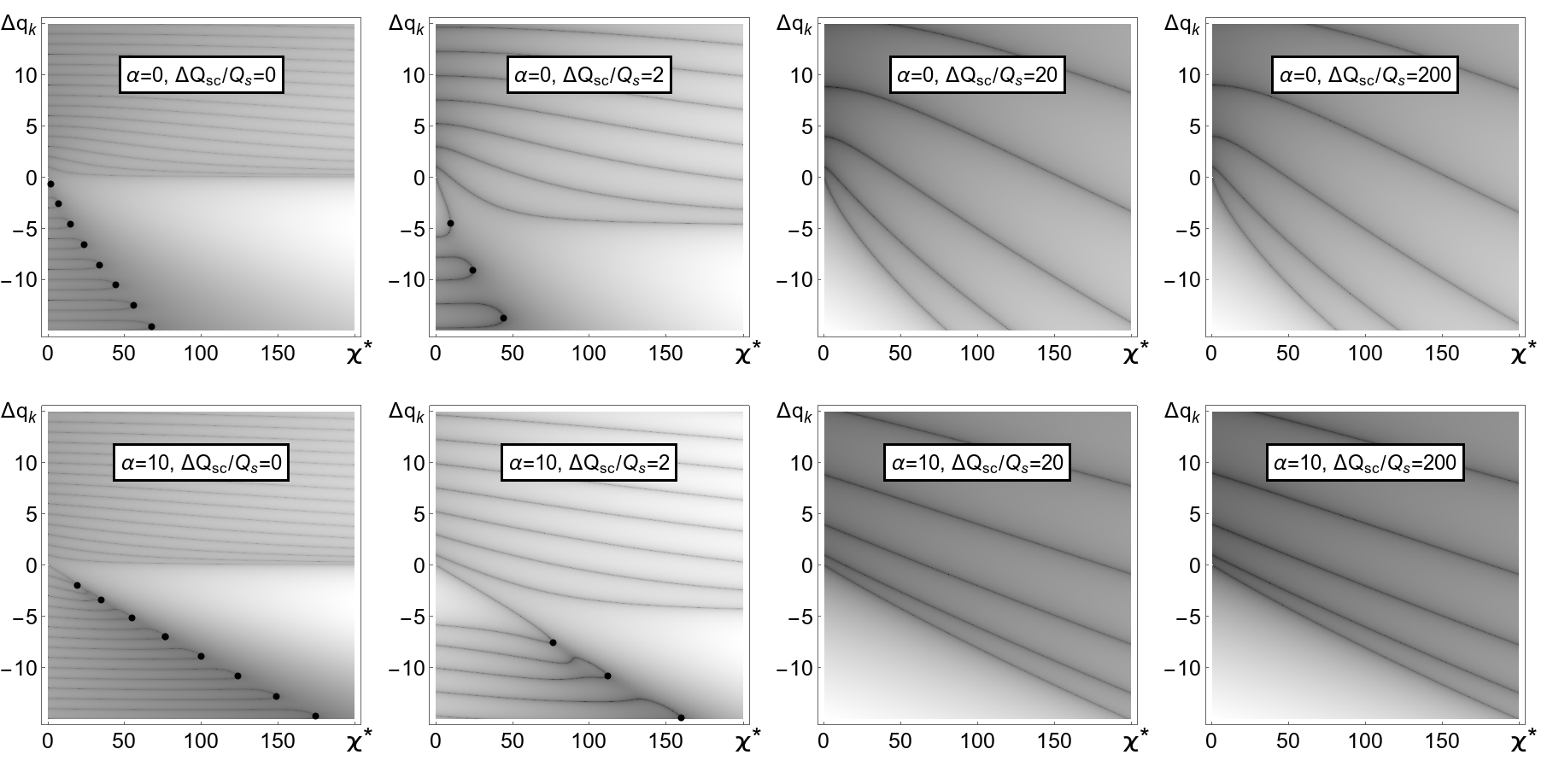}
\caption{\label{fig:ABExpWakeSpectra}
	Spectra for the ABS model and constant (top row) or exponential (bottom 
row)
	wakes.
	Different columns correspond to different values of the space 
	charge parameter $\Qsc/\Qs=0,2,20,200$.
	The TMCI thresholds are shown by black points.
	}
\end{figure*}
%------------------------------------------------------------------------------%

%==============================================================================%
%------------------------------------------------------------------------------%
%------------------------------------------------------------------------------%
%------------------------------------------------------------------------------%
%==============================================================================%
\section{\label{sec:NegWakes}Negative wakes}

Without wakes, all the modes are divided into two groups, with positive and 
negative tune shifts $\Qx$; at growing SC, the former tend to zero, while the 
latter tend to the SC tune shift. It is convenient to keep the same mode number 
when the wake, when introduced, tunes it up or down. Therefore, here-below a 
number of a particular mode for non-zero wake is defined as such of that 
zero-wake mode, which corresponds to the given mode when the wake amplitude 
continuously goes to zero. This mode number definition is unambiguous provided 
that the given mode never coupled with its neighbors at smaller wake amplitudes, 
which is sufficient for our purposes. Following that, we distinguish all the 
modes at any wake between {\it positive and negative modes}, or {\it positive or 
negative parts of the spectrum}, which is unambiguous provided that there is no 
coupling between the positive and negative modes, which is always the case 
below. {\it The reader should not be confused by the fact that wakes may shift 
tunes of some positive modes to negative values; we still refer to such modes as 
positive.}              

%Even when a wake will be introduced, we will still refer to modes in a spectrum
%by an integer index $k$ associated with the no-wake solution.
%Modes with $k>0$ ($<0$) will be referred as belonging to  {\it the positivepart
%of spectrum} ({\it the negative part of spectrum}).

%------------------------------------------------------------------------------%
In this section, constant-sign wakes are considered for the ABS, as well as for 
the SSC problems;
as it follows from the Maxwell equations, this sign can only be negative
\cite{chao1993physics}.
We limit ourselves here by delta-functional, exponential and resistive wall
wakes.
Sometimes, the tunes at a given wake and SC are convenient to present in units 
of the first eigenvalue at the same SC and zero wake.
For that sake we will modify the intensity parameter $\kappa$ such that it will
be measured in the units of
\begin{equation}
	b^2/[\Qef(0)\,a^2]\quad\text{and}\quad
	b^2/[2\,\Qef(0)\,a^2],
\end{equation}
for the SSCSW and the SSCHP models respectively.
We hope that the factor of 2 difference here will not confuse the reader; it is 
dictated by our 
choice to have same formulas for the first eigenvalue for the two SSC cases at 
zero wake. 

Dividing $\kappa$ by the first zero-wake eigenvalue at the given SC
parameter, we define a {\it normalized} intensity parameter
\begin{equation}
	\kappa^* = \kappa \Big/ \left[-\frac{\Qsc}{2\,\Qs} + \
	\sqrt{\left(\frac{\Qsc}{2\,\Qs}\right)^2 + 1}\right].
\end{equation}
When $\kappa^*$ is used instead of $\kappa$, the positive part of the
spectrum shows scale invariance for large values of $\Qsc$,
Eq.~(\ref{math:SSC}).
In addition, in order to describe the instability we will introduce a parameter
which is combining the intensity and the wake amplitude, respectively
\begin{equation}
\chi   = \frac{\kappa  \,W_0\tb}{\Qs}
\qquad\text{and}\qquad
\chi^* = \frac{\kappa^*\,W_0\tb}{\Qs}.
\end{equation}

%==============================================================================%
%------------------------------------------------------------------------------%
%==============================================================================%
\subsection{\label{sec:DeltaWake}Delta wake}

%------------------------------------------------------------------------------%
Our first negative-wake example is one of image charges, the delta
wake,
\begin{equation}
	W(\tau) = - W_0\,\delta(\tau).
\end{equation}
In this case the airbag model allows analytical solution
\begin{equation}
	\frac{\Qx}{\Qs} = -\frac{\Qsc+\kappa\,W_0}{2\,\Qs} \pm \
	\sqrt{\left(\frac{\Qsc-\kappa\,W_0}{2\,\Qs}\right)^2 + k^2},
\end{equation}
for $k = \pm 1,\pm 2,\ldots$ and the zeroth mode $\Delta Q_0/\Qs =-\kappa\,W_0$.
The system is stable for any values of wake amplitude and SC, 
Fig.~\ref{fig:DeltaWake} shows normalized tune shift as a function of wake 
amplitude for different values of SC.

%------------------------------------------------------------------------------%
For the SSC models, matrix elements are given by
\begin{equation}
	\mW =
	-W_0\,\int_\text{SB} \rho^2(\tau)\,Y_l(\tau)\,Y_m(\tau)\,\dd\tau =
	-W_0\,\frac{\delta_{lm}}{2},
\end{equation}
yielding the eigenvalues
\begin{equation}
	\qx = \nu_k + \kappa\,\widehat{\text{W}}_{kk} = \nu_k - 
	\frac{\kappa\,W_0}{2}.
\end{equation}

%------------------------------------------------------------------------------%
Beam stability for the delta wake is a consequence of that the system is 
Hamiltonian: the delta wake can be taken into account with a term proportional 
to a 
double sum $\Sigma_{lm} x_l x_m \delta (\tau_l - \tau_m)$ in the total 
Hamiltonian, where $x_l$ and $x_m$ are offsets of individual particles and 
$\tau_l$, $\tau_m$ are their positions inside the bunch.

%==============================================================================%
%------------------------------------------------------------------------------%
%==============================================================================%
\subsection{\label{sec:ExpWake}Exponential and constant wakes}

%------------------------------------------------------------------------------%
As the next example, let's consider exponential wakes
\begin{equation}
	W(\tau) = -W_0\,e^{\alpha\,\tau},
\end{equation}
including a constant wake, $\alpha=0$.
Behavior of TMCI with these wakes was considered for ABS by M.~Blaskiewicz and
for HP and SW cases by V.~Balbekov
\cite{balbekov2016dependence,balbekov2016tmci,balbekov2017transverse,
balbekov2017threshold}.
Here we summarize some of these results and make a comparison between 
models.

\begin{figure}[t!]
\includegraphics[width=0.9\linewidth]{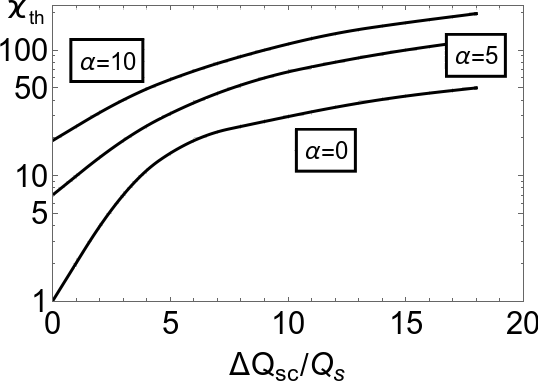}
\caption{\label{fig:ABExpWakeSummary}
	TMCI threshold as a function of space charge for the ABS model with the 
	exponential wakes (reproduction of Ref.~\cite{blaskiewicz1998fast}).
	}
\end{figure}

%------------------------------------------------------------------------------%
Fig.~\ref{fig:ABExpWakeSpectra} shows normalized coherent tune shifts for
constant and exponential ($\alpha=10$) wakes for the ABS model.
As one can see, TMCI may occur only at the negative part of the spectrum and the
thresholds monotonically increase with $\Qsc$.
Last two plots in each row look identical to each other, thus showing the SSC 
scaling.
Behavior of the TMCI thresholds is summarized in Fig.~\ref{fig:ABExpWakeSummary}
which we reproduce, albeit by another method, after M.~Blaskiewicz
\cite{blaskiewicz1998fast}.

%------------------------------------------------------------------------------%
%------------------------------------------------------------------------------%

%------------------------------------------------------------------------------%
\begin{figure}[b!]
\includegraphics[width=\linewidth]{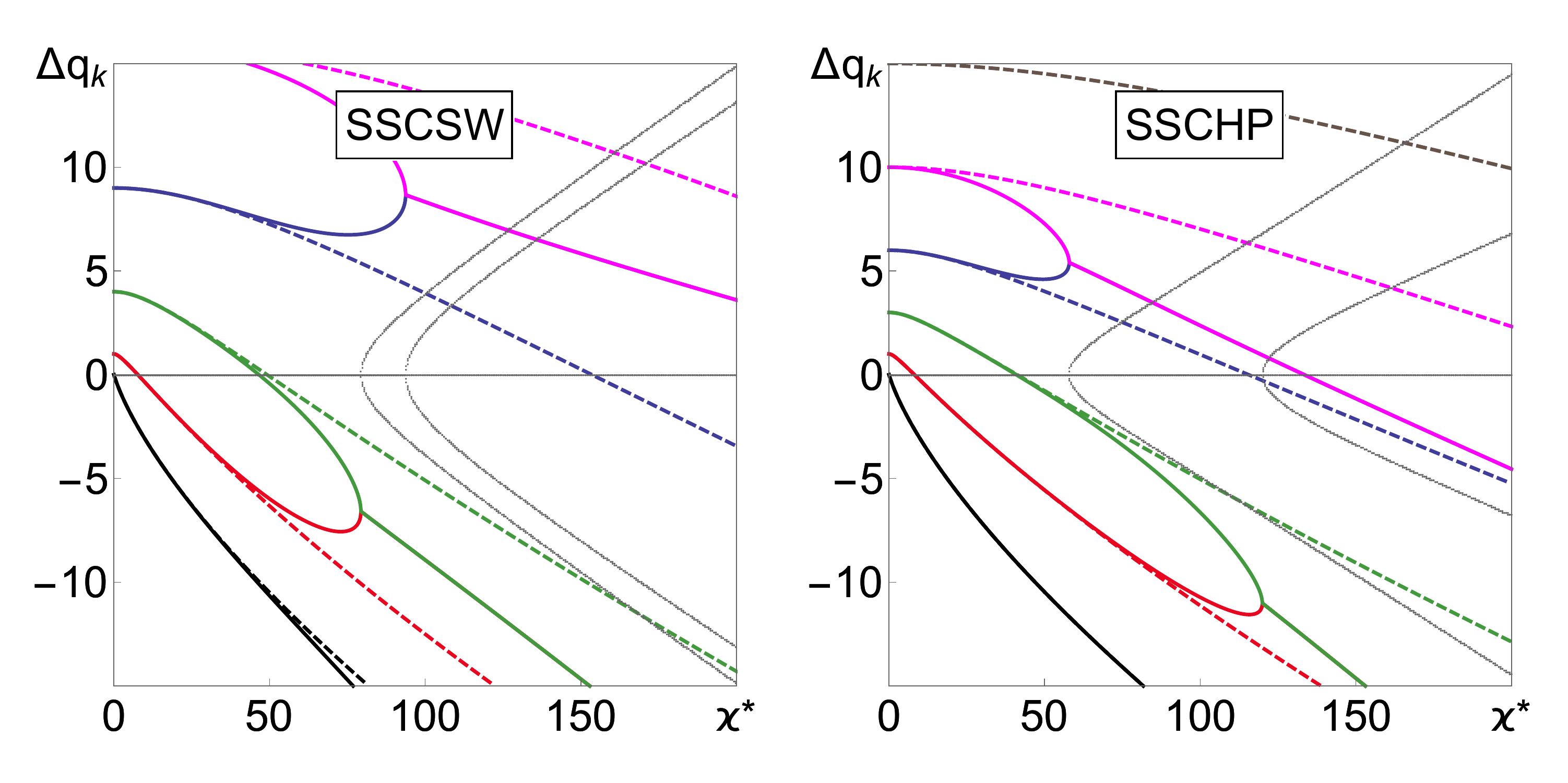}
\caption{\label{fig:SSCConstWakeSpectra}
	Real (shown in color) and imaginary (gray colors) parts of the spectra
	for SSCSW and SSCHP with a constant wake.
	Solid and dashed lines are obtained using 5 or 50 modes respectively.
	For sufficiently large number of modes and sufficiently good accuracy of
	the matrix element computations, TMCI vanishes.
	}
\end{figure}
%------------------------------------------------------------------------------%

%------------------------------------------------------------------------------%
Increase of the TMCI thresholds with the space charge, first demonstrated in 
Ref.~\cite{blaskiewicz1998fast}, soon after that was explained in 
Ref.~\cite{ng1999stability}:
while the wake deflects down mostly the zeroth mode, and to a lesser extent does
that for the mode $\Delta Q_{-1}$, the space charge does exactly the opposite, 
deflecting down the mode $\Delta Q_{-1}$ and keeping the tune of the mode $0$ 
untouched;
thus, the mode crossing either happens at higher wakes, or does not happen at 
all.
However, it was later speculated in Ref.~\cite{burov2009head} that the increase
of the threshold with the space charge may be non-monotonic, that at 
sufficiently high SC tune shift, the TMCI threshold may start to go down.
A reason for that speculation was derived from the unlimited reduction of the
mode separation with the space charge;
when the neighbor modes are closer and closer, it should take less and less to 
couple them.
This speculation was apparently confirmed within the SSC approximation by 
computation of the TMCI threshold versus the SC tune shift with constant and 
resistive-wall wakes, where the non-monotonic behavior of the instability 
threshold was observed.
However, those computations of Ref.~\cite{burov2009head} were made with not so
high number of modes and with limited accuracy of the matrix element 
computation, so the formulated hypothesis remained open.
This problem was recently addressed by 
V.~Balbekov~\cite{balbekov2017transverse,balbekov2017threshold}.
Specifically, it has been shown by him, that the TMCI threshold computed for 
the exponential wakes unlimitedly grows with the number of the basis functions 
taken into account.
In other words, for such wakes at SSC case there is no TMCI at all, the
instability may happen only with comparable SC and wake-driven tune shifts, as 
in Fig.~\ref{fig:ABExpWakeSummary}.
Here we confirm that with our results presented in
Fig.~\ref{fig:SSCConstWakeSpectra}.
Solid lines show real and imaginary parts of the spectra for SSCSW and SSCHP 
cases for a constant wake and truncation at 5 harmonics.
Dashed lines are the same spectra computed with 50 basis harmonics.
As one can see, no instability observed in the considered range of $\kappa$ when
number of harmonics is large enough.
Note agreement between the ABS model (constant wake and large $\Qsc$, top right 
plot in Fig.~\ref{fig:ABExpWakeSpectra}) and the SSCSW model (constant wake for 
large number of modes, left plot in Fig.~\ref{fig:SSCConstWakeSpectra}).

\begin{figure}[t!]
\includegraphics[width=\linewidth]{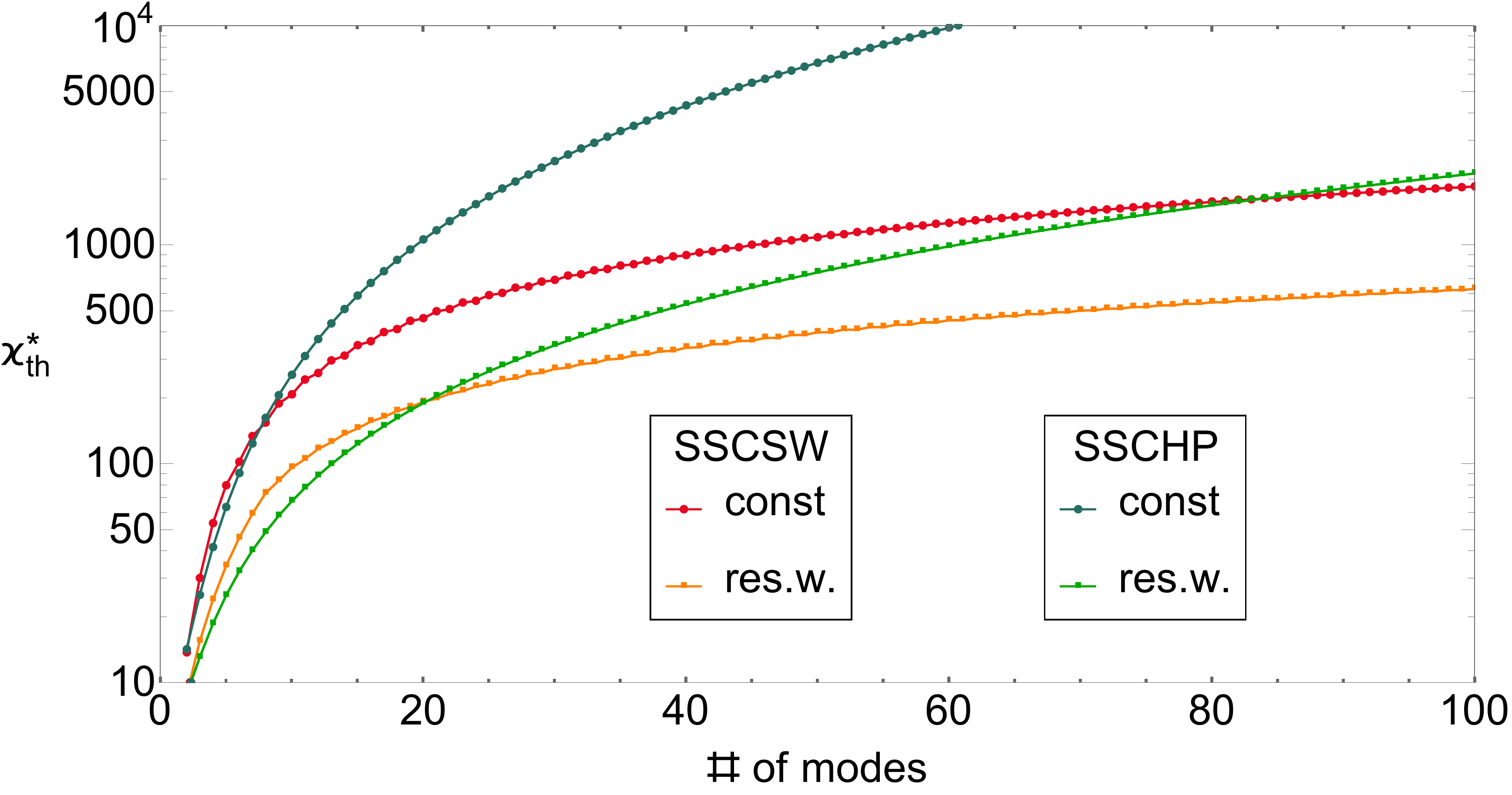}
\caption{\label{fig:SSCNegWakeConv}
	TMCI threshold as a function of number of modes for constant and
	resistive wall wakes (SSCSW and SSCHP models).
	The unlimited growth with the truncation parameter means that there is
	no TMCI.
	}
\end{figure}

%------------------------------------------------------------------------------%
Figure~\ref{fig:SSCNegWakeConv} shows growth of the instability thresholds with
the number of harmonics in analysis for both models (SSCSW and SSCHP) with the
constant wake.
For all the cases, the thresholds monotonically increase with the mode
truncation parameter, thus demonstrating that the beam is stable against TMCI.

When the instability threshold unlimitedly increases with SC, the TMCI is 
referred in this paper as {\it vanishing}.

%------------------------------------------------------------------------------%
%------------------------------------------------------------------------------%

%------------------------------------------------------------------------------%
\begin{figure}[h!]
\includegraphics[width=\linewidth]{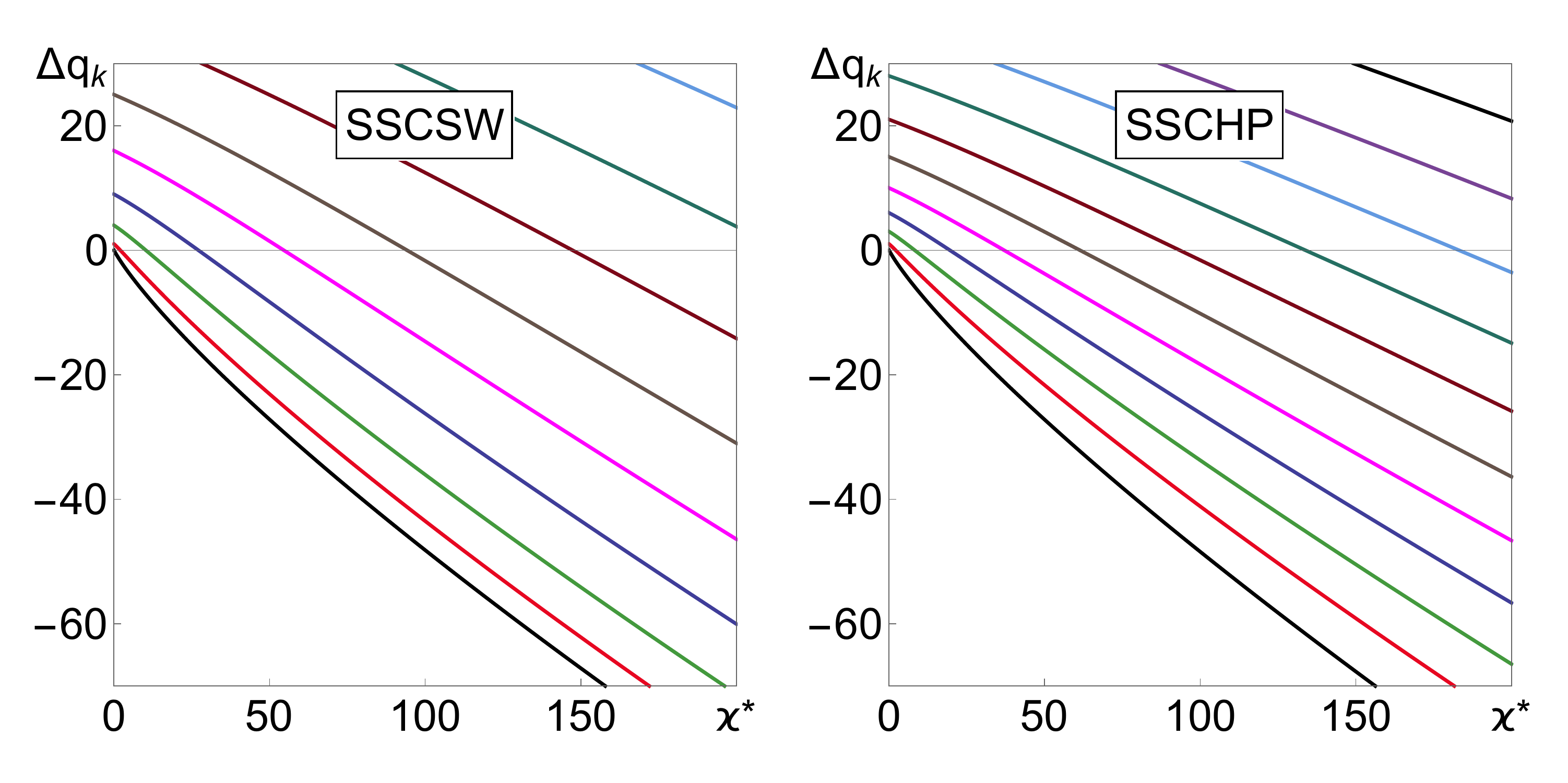}
\caption{\label{fig:SSCResWakeSpectra}
	Spectra for SSCSW and SSCHP models with resistive wall
	wake, truncated at 50 modes. No TMCI.
	}
\end{figure}
%------------------------------------------------------------------------------%

%------------------------------------------------------------------------------%
\begin{figure*}[p]
\includegraphics[width=\linewidth]{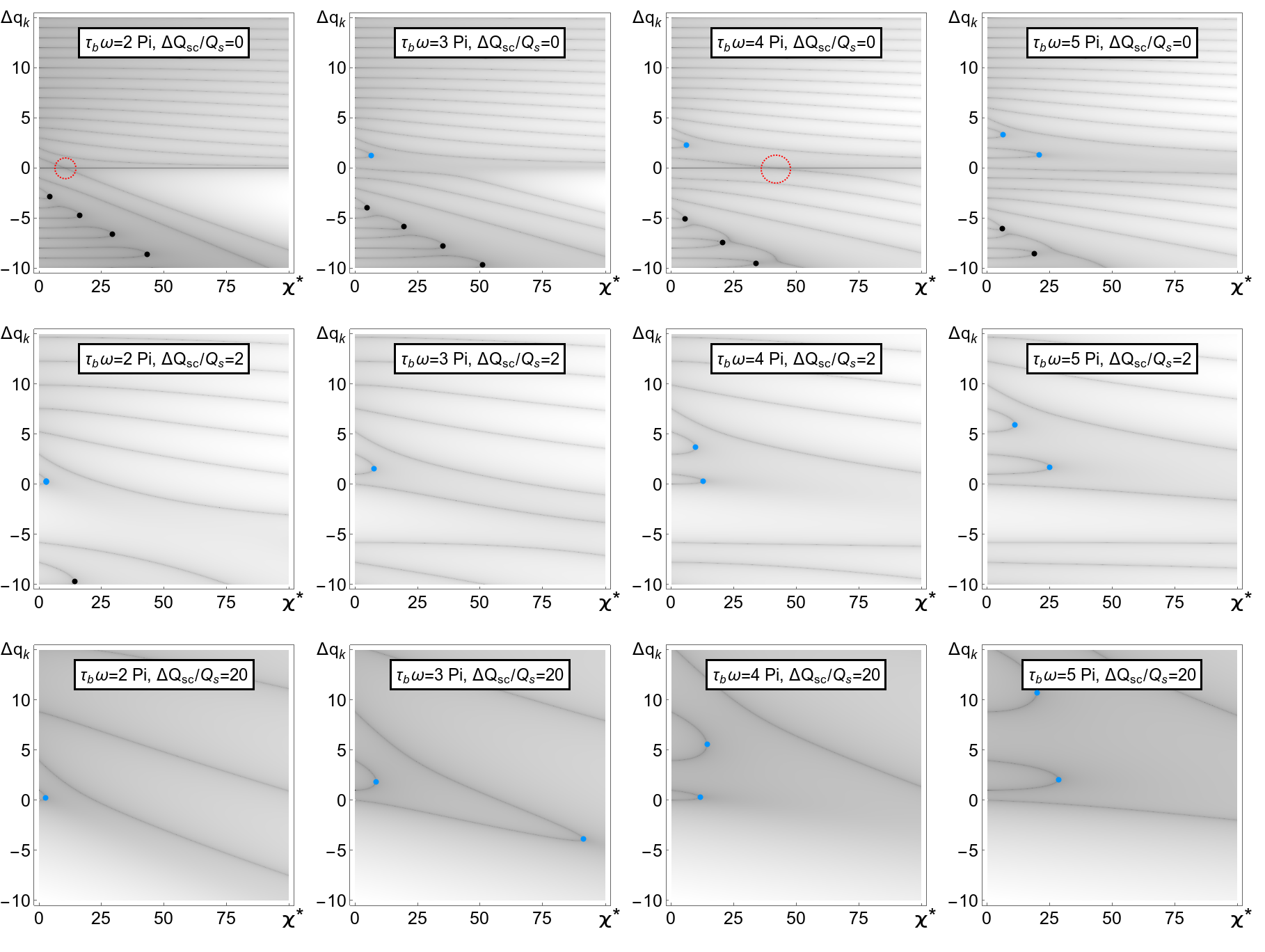}
\includegraphics[width=\linewidth]{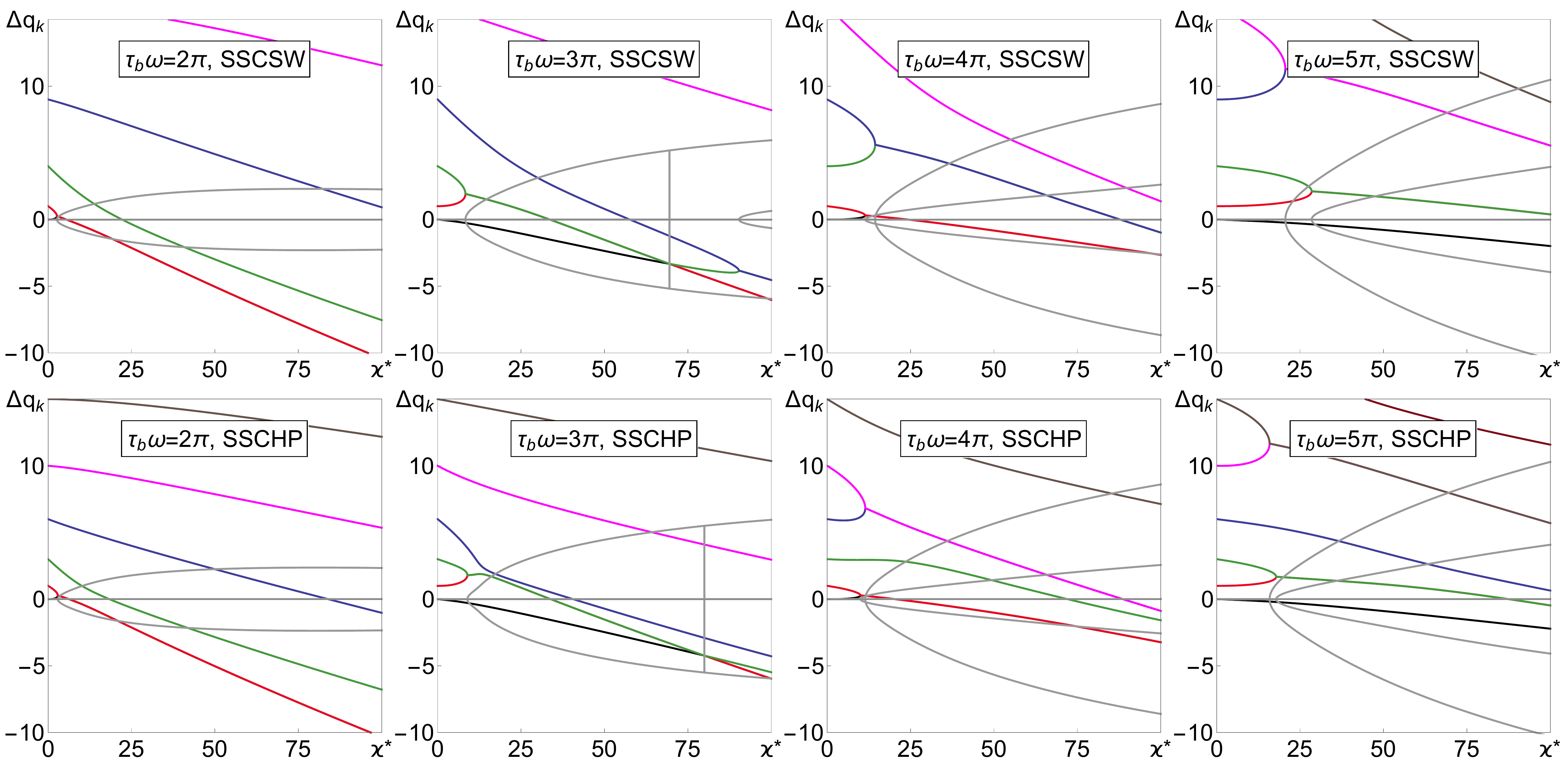}
\caption{\label{fig:CosWakeSpectra}
	Bunch spectra for the cosine wake.
	The top three rows show the ABS spectra for various values of SC
	(0, 2, 20).
	The vanishing TMCI thresholds are shown with black points and 
	non-vanishing with blue.
	The red dashed circles indicate {\it mode crossing}, when modes do not 
	couple.
	The last two rows are for the SSCSW and SSCHP models;
	the real and imaginary parts of $\qx$ are shown in colors and in 
	gray respectively.
	The columns correspond to different values of $\omega\,\tb$
	($2\,\pi,3\,\pi,4\,\pi,5\,\pi$).
	}
\end{figure*}
%------------------------------------------------------------------------------%

%==============================================================================%
%------------------------------------------------------------------------------%
%==============================================================================%
\subsection{\label{sec:ResWake}Resistive wall wake}

%------------------------------------------------------------------------------%
The SSC models can also be considered with the resistive wall wake function
\begin{equation}
	W(\tau) =
		-W_0/\sqrt{|\tau|}.
\end{equation}
Similar to the exponential and constant wakes, the thresholds grow 
unlimitedly with the number of modes taken into account, as it can be seen in 
Fig.~\ref{fig:SSCNegWakeConv}.
Spectra for both SSCSW and SSCHP cases are shown in 
Fig~\ref{fig:SSCResWakeSpectra}.

%==============================================================================%
%------------------------------------------------------------------------------%
%------------------------------------------------------------------------------%
%------------------------------------------------------------------------------%
%==============================================================================%

\section{\label{sec:OscWakes}Oscillating wakes}

%------------------------------------------------------------------------------%
As shown above, the instability for negative wakes takes place only when
wake and space charge tune shifts are comparable, i.e. the TMCI is of the {\it 
vanishing} type.
At positive (non-physical) wakes, the instability threshold monotonically 
decreases with 
the space charge;
the wake moves 0-th tune up, while the space charge deflects the tune of the 
above 1-st mode down, so SC helps the two modes to meet each other.
However, the positive wake is non-physical, it is forbidden by Maxwell's
equations~\cite{chao1993physics}.
What is possible, is a combination of positive and negative wakes, i.e. 
oscillatory wakes.
It can be expected that, at least for some of them, the instability threshold
decreases with the SC, similar to the positive wakes.
In fact, this was already shown for the cosine wake in
Ref.~\cite{burov2009head}:
contrary to negative, the cosine wake may deflect down a certain mode more than 
its lower neighbor, thus making modes crossing and possibly couple.
Since SC reduces the mode separation, in this case, the instability threshold 
goes down with an increase of the SC tune shift.

%------------------------------------------------------------------------------%
In this section, we consider oscillating wakes of two types:
cosine and sine with variable decay rates.
All of them can be treated within the ABS model, as suggested in 
Ref.~\cite{blaskiewicz1998fast}.
For comparison, results of the complimentary SSC approximation are also 
provided below.

%==============================================================================%
%------------------------------------------------------------------------------%
%==============================================================================%
\subsection{\label{sec:CosWake}Cosine wake}

\begin{figure}[t!]
\includegraphics[width=0.95\linewidth]{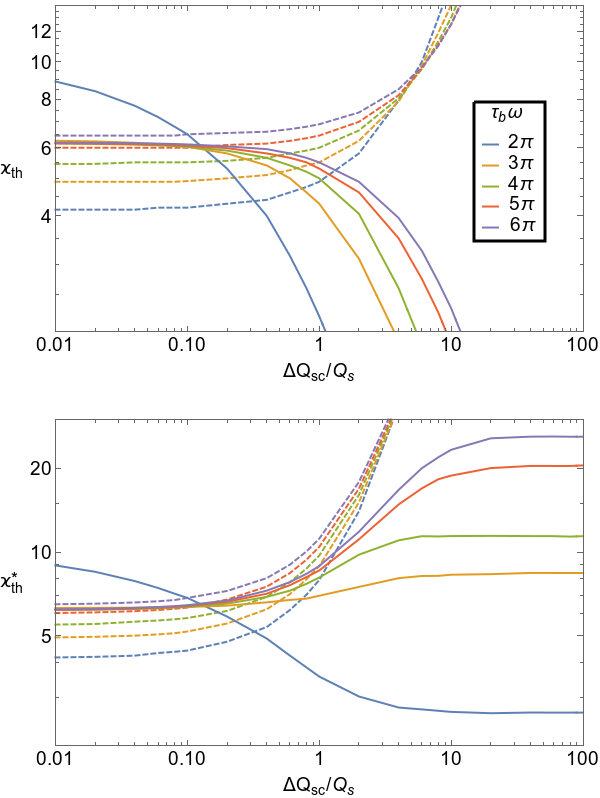}
% {FIGURE_ABCosWakeSummary.pdf}
\caption{\label{fig:ABCosWakeSummary}
	TMCI threshold as a function of space charge for the ABS model and a
	cosine wake.
	The dashed and solid lines correspond to the vanishing and non-vanishing
	TMCIs.
	The top figure shows thresholds in regular units.
	The bottom figure shows the same but using normalized eigenvalues: at 
every 
	SC parameter,
	the eigenvalues are related to one of the first mode at zero wake. 
	}
\end{figure}

%------------------------------------------------------------------------------%
For cosine wakes
\begin{equation}
	W(\tau) =
		-W_0 \cos(\omega\,\tau),
\end{equation}
the normalized spectra for different values of the SC tune shift and the wake 
phase 
advance $\omega\,\tb$ are shown in Fig.~\ref{fig:CosWakeSpectra}.
It shows that the situation is more diverse here than for the negative 
wakes: in addition to instabilities in the negative part of the spectra 
(indicated with black points again), there is a mode coupling in the positive 
parts of the spectra (blue points) as well.
Figure~\ref{fig:ABCosWakeSummary} summarizes the behavior of the lowest mode 
coupling thresholds for the modes with negative and positive indexes.
While TMCIs in the negative part of the spectra vanish, as it was for the case 
of
constant-sign wakes, TMCI thresholds in the positive part of the spectra are 
non-monotonic, as it was speculated in Ref.~\cite{burov2009head}; being growing 
at sufficiently small SC parameters, the thresholds become  
inversely proportional to the SC tune shift at its higher values, i.e. TMCI does 
not vanish.
On the bottom plot, which repeats the top one in normalized units, one can see 
the stabilization of the non-vanishing TMCI threshold when modes, which 
cause the instability, reach the SSC limit.

\begin{figure}[h!]
\includegraphics[width=0.95\linewidth]{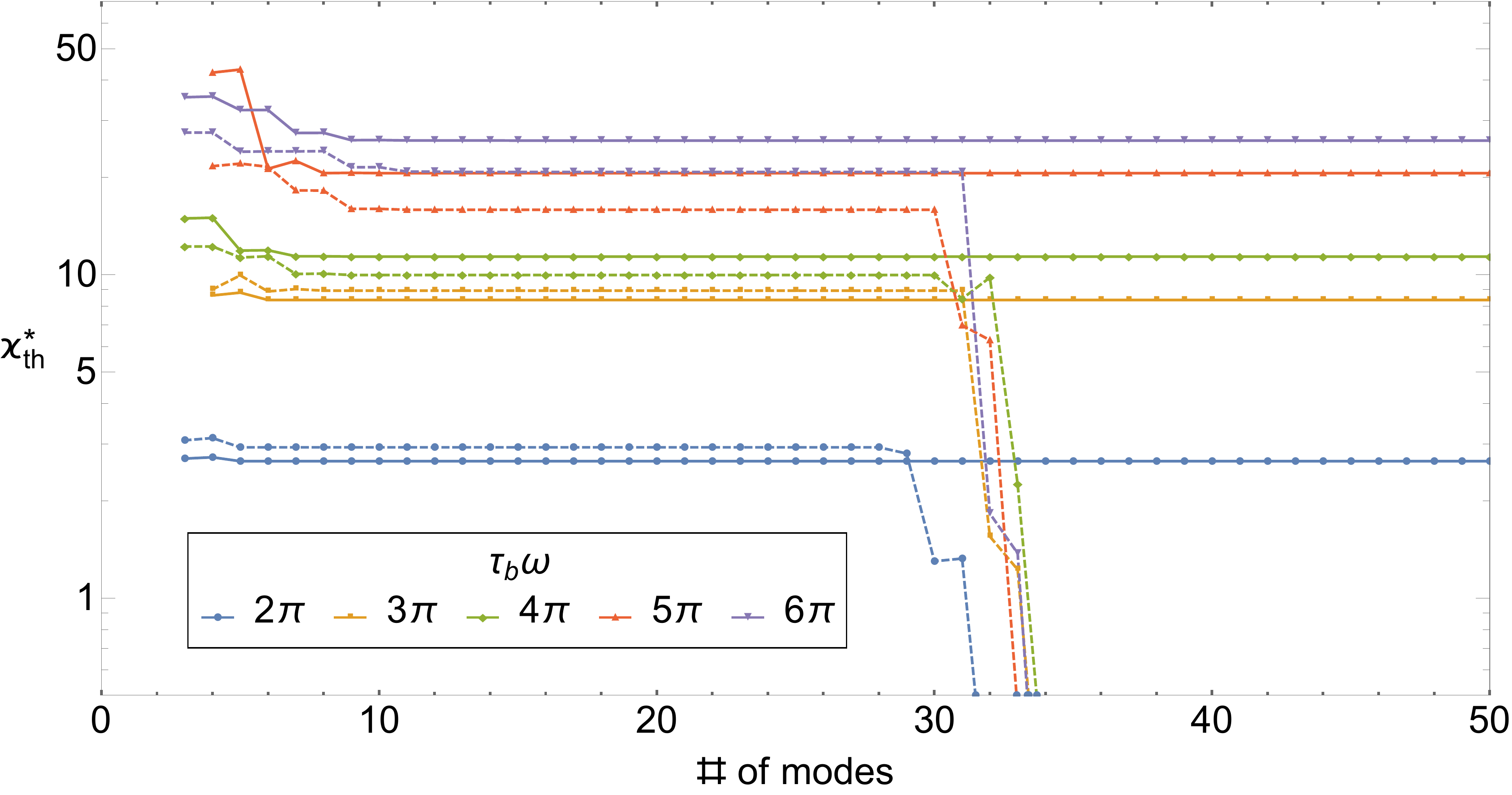}
\caption{\label{fig:SSCCosWakeConv}
	Instability threshold as a function of the number of modes taken into 
	account at the SSC for the cosine wake.
	SSCSW and SSCHP models are shown using solid and dashed lines 
	respectively.
	For the SSCHP model when the number of modes is $\approx 30$,
	there is a fast decrease of $\kt$ reflecting the numerical error;
	such a numerical error was not observed for the SSCSW case up to 100 
	modes.
	}
\end{figure}

%------------------------------------------------------------------------------%
At high SC parameter, all the models give similar results, while the SSCSW model
fully coincides with the ABS (compare third and fourth rows in 
Fig.~\ref{fig:CosWakeSpectra}).
In contrast to the case of negative wakes, the threshold computed for the SSC
quickly converges with the number of modes taken into account.
Examples of convergence of $\kt$ are presented in Fig.~\ref{fig:SSCCosWakeConv}.
At the same time, when the number of modes becomes too large (which is about 30 
in our case), the SSCHP model shows a pure numerical instability: the 
threshold computations become unstable and its erratically computed value drops 
to 0.

%------------------------------------------------------------------------------%
%------------------------------------------------------------------------------%

%------------------------------------------------------------------------------%
%------------------------------------------------------------------------------%

\begin{figure}[h!]
\includegraphics[width=\linewidth]{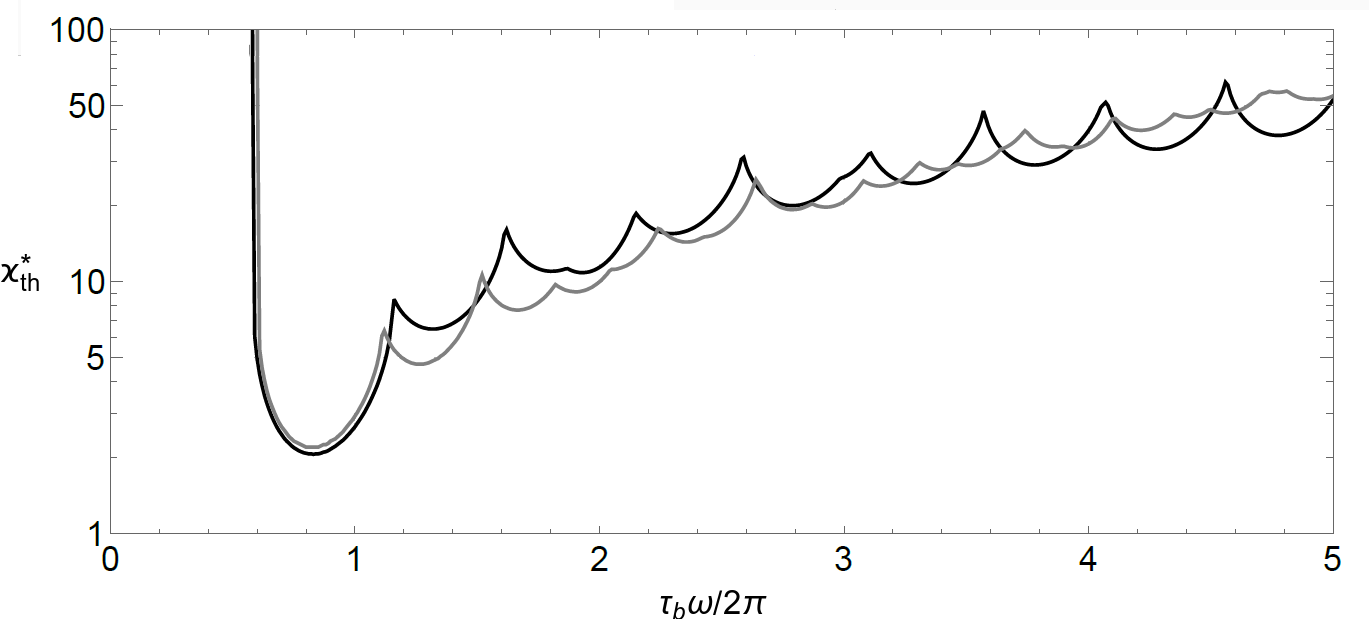}
\caption{\label{fig:SSCCosWakeOmega}
	TMCI threshold as a function of wake phase advance $\omega\,\tb$,
	for SSCSW (black) and SSCHP (gray) models with cosine wake.
	The vertical lines reflect an absolute threshold with respect to 
	$\omega\,\tb$.
	}
\end{figure}

%------------------------------------------------------------------------------%
In order to summarize the behavior of $\kt$ for SSC and the cosine wake, we 
considered it as a function of $\omega\,\tb$ for both models, see
Fig.~\ref{fig:SSCCosWakeOmega}.
As expected, TMCI has a threshold with respect to $\omega\,\tb \approx \pi$:
for smaller values of the phase advance, the wake is effectively negative.

%------------------------------------------------------------------------------%
%------------------------------------------------------------------------------%

%------------------------------------------------------------------------------%
\begin{figure*}[p]
\includegraphics[width=\linewidth]{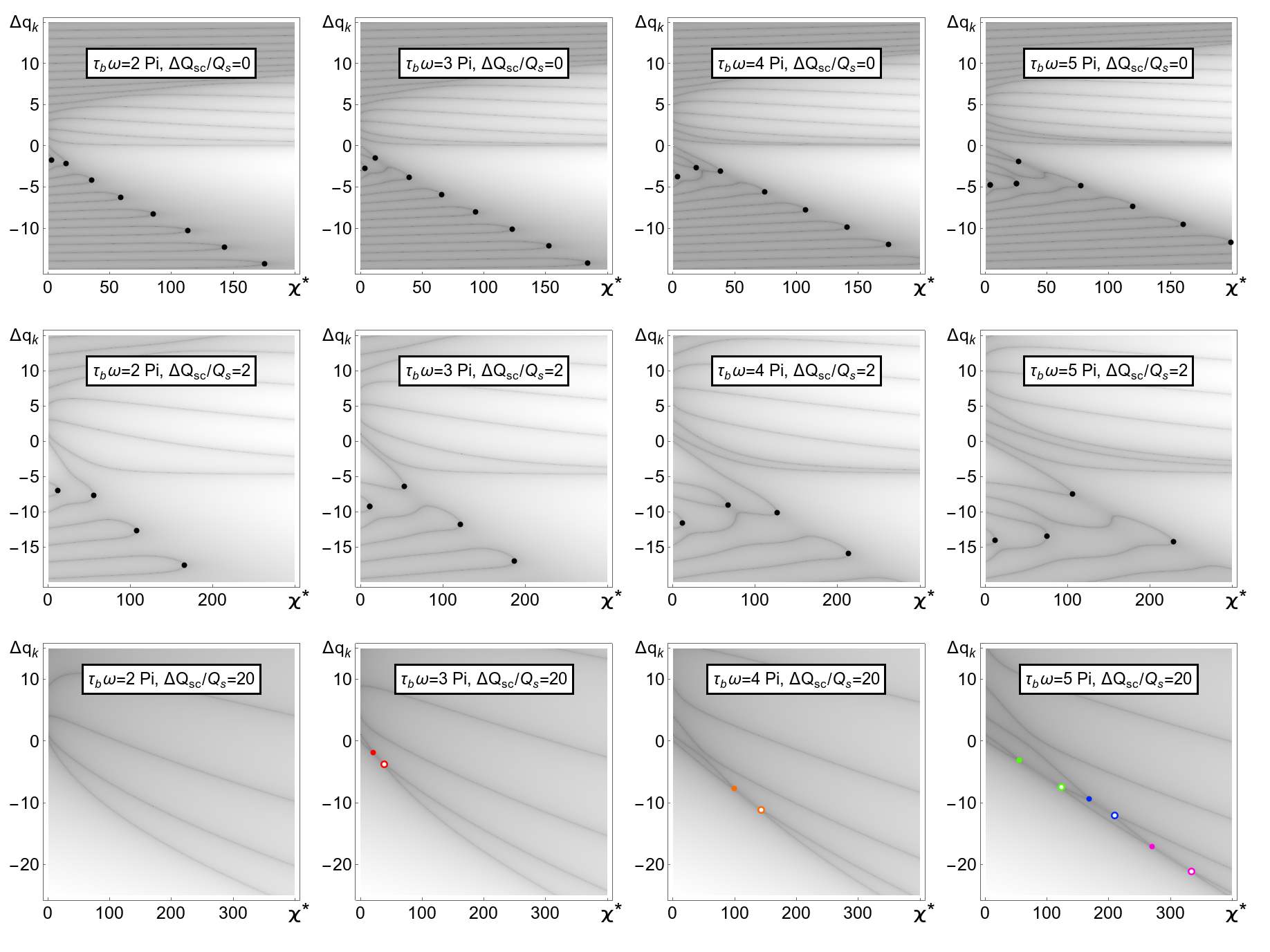}
\includegraphics[width=\linewidth]{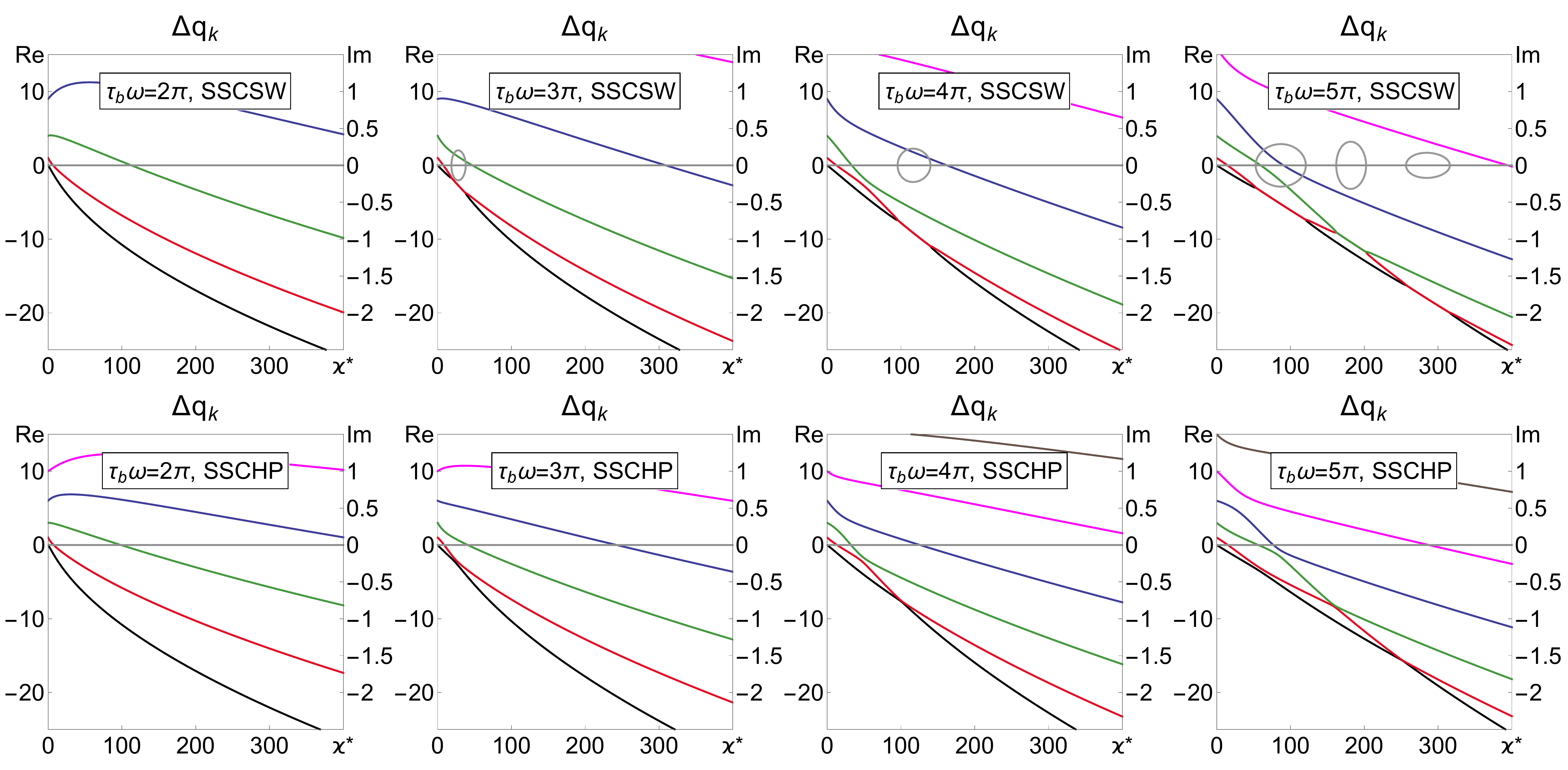}
\caption{\label{fig:SinWakeSpectra}
	Bunch spectra for the sine wake.
	The top three rows show the ABS spectra for various values of SC 
	(0, 2, 20).
	The vanishing TMCI thresholds are shown with black and non-vanishing 
	using points in colors (subsequent decoupling for each TMCI is shown by 
	the point of the same color with an annulus).
	The last two rows are for the SSCSW and SSCHP models;
	the real and imaginary parts of $\qx$ are shown in colors and in 
	gray respectively (note the different scaling for $\Delta_k$ and 
	$\Gamma_k$).
	The columns correspond to the different values of $\omega\,\tb$
	($2\,\pi,3\,\pi,4\,\pi,5\,\pi$).
	}
\end{figure*}
%------------------------------------------------------------------------------%

%==============================================================================%
%------------------------------------------------------------------------------%
%==============================================================================%
\subsection{\label{sec:SinWake}Sine wake}

\begin{figure}[t!]
\includegraphics[width=\linewidth]{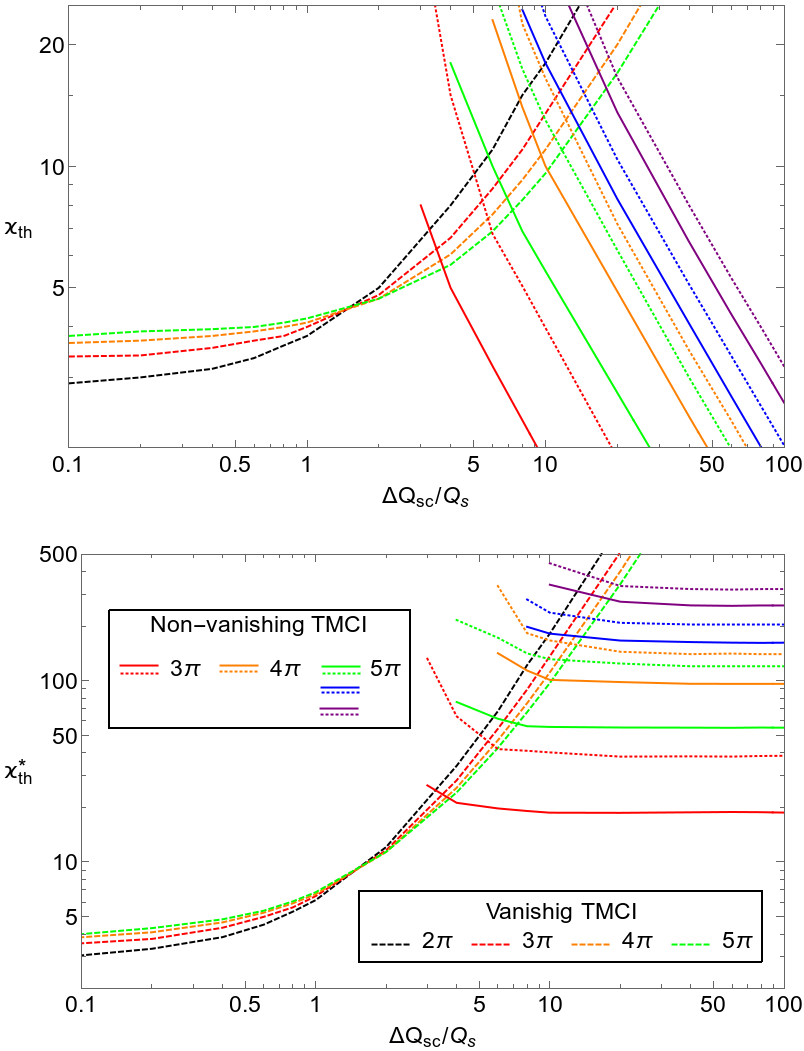}
\caption{\label{fig:SinWakeABSummary}
	TMCI threshold in the usual (top) and normalized (bottom) units as a 
function of space charge for ABS
	model and sine wake.
	The dashed lines correspond to the vanishing TMCI in a negative part of 
	the spectra.
	The solid and dotted lines represent mode couplings and
	decouplings in the positive part of the spectra.
	}
\end{figure}

%------------------------------------------------------------------------------%
The sine wake
\begin{equation}
	W(\tau) = W_0\,\sin(\omega\,\tau),
\end{equation}
is considered for the same three models, ABS, SSCSW and SSCHP;
the resulting spectra are shown in Fig.~\ref{fig:SinWakeSpectra}.
The behavior of the negative modes looks similar to the case of the cosine wake.
In contrast, when the SC is zero, there is no instability for the positive 
modes, independently of the wake phase advance 
$\omega \tb$.
However, when the SC is increased, a cascade of mode couplings and decouplings 
appears.
When $\omega\,\tb=2\,\pi$, the threshold increases with the SC parameter, 
disappearing at the SSC case. 
The cases of $\omega\,\tb=3\,\pi,4\,\pi$ show a single mode coupling followed 
by decoupling;
the case $\omega\,\tb=5\,\pi$ has a cascade of three couplings with subsequent 
decouplings.
For wakes with even more oscillations per bunch, more complicated 
coupling-decoupling cascades were observed (not shown in the figure).
Figure~\ref{fig:SinWakeABSummary} summarizes the behavior of the lowest TMCI 
thresholds for both negative and positive modes.
While all instabilities in the positive part of the spectrum decouple at higher 
values of $\kappa$, the values of these coupling and decoupling intensities go 
down inversely proportional to the SC parameter. 

%------------------------------------------------------------------------------%
%------------------------------------------------------------------------------%

%------------------------------------------------------------------------------%
\begin{figure}[b!]
\includegraphics[width=\linewidth]{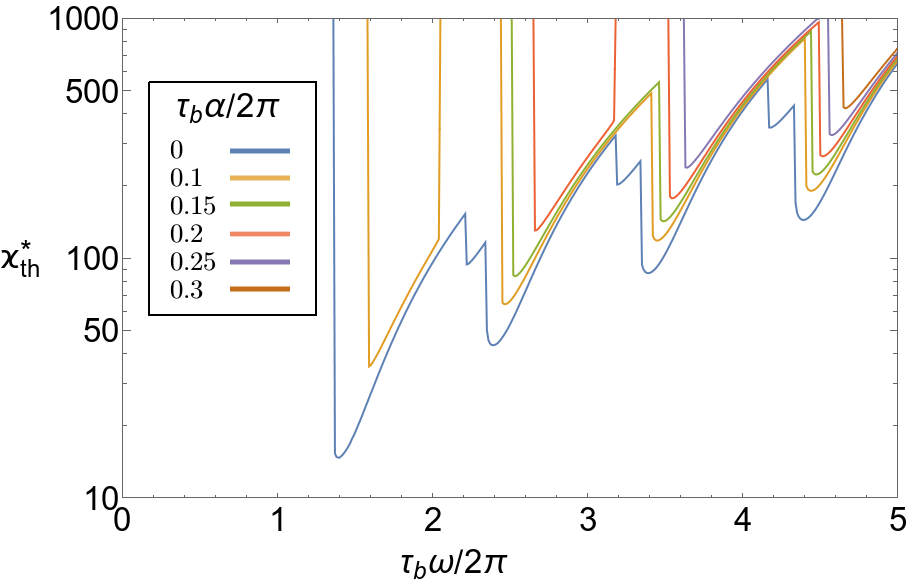}
\caption{\label{fig:SSCSinWakeOmega}
	TMCI threshold for SSCSW model with broadband wake as a 
	function of wake phase advance $\omega\,\tb$.
	The different curves correspond to the different values of the wake 
	decay rate, $\alpha\,\tb$.
	The blue curve with $\alpha\,\tb = 0$ is for the sine wake.
	The vertical lines reflect an absolute threshold with respect to 
	$\omega\,\tb$.
	}
\end{figure}
%------------------------------------------------------------------------------%

%------------------------------------------------------------------------------%
Figure~\ref{fig:SinWakeSpectra} shows that the SSCSW results agree with the ABS 
ones when the SC parameter is high enough (as they have to), while the SSCHP 
gives something unexpectedly different.
Although the spectra of the SSCHP may look similar to both ABS and SSCSW 
models, they are different: the SSCHP modes never couple.
Instead, they either simply cross or approach-divert with respect to each other
without coupling.
For this specific case there is an unknown reason forbidding the instability.
This very unique property of the spectra is structurally 
unstable: the perturbation of $\nu_k^{\text{hp}}$ leads to 
the coupling-decoupling with instability, instead of mode crossing or 
approach-diverting.

%------------------------------------------------------------------------------%
The behavior of the non-vanishing TMCI as a function of $\omega\,\tb$ for SSCSW 
is shown in Fig.~\ref{fig:SSCSinWakeOmega}.
Similar to the cosine wake, the instability is impossible when
$\omega\,\tb \leq 2\,\pi$;
otherwise, it happens and its threshold is a non-monotonic function of the wake 
phase advance $\omega\,\tb$.

%==============================================================================%
%------------------------------------------------------------------------------%
%==============================================================================%
\subsection{\label{sec:ROsc}Broadband wake}

%------------------------------------------------------------------------------%
Finally, we consider how the spectrum for the sine wake is modified when a 
certain decay is added:
\begin{equation}
	W = W_0\,\sin(\omega\,\tau)\,e^{\alpha\,\tau};
\end{equation}
conventionally, this wake function is referred as broadband.

The result for the SSCSW model is shown in Fig.~\ref{fig:SSCSinWakeOmega}.
There is no surprise that the instability threshold increases with $\alpha$ for 
fixed $\omega$.
When a certain threshold with respect to the decay parameter $\alpha$ is 
reached, the instability vanishes: strong enough suppression of the oscillating 
part by the 
exponential effectively makes the wake negative.
Thus, the TMCI does vanishes in certain zones of the wake decay and oscillation
parameters $\alpha \tb$ and $\omega \tb$.

%------------------------------------------------------------------------------%
The ABS model confirms this result: for fixed $\omega$, when a threshold with 
respect to $\alpha$ is reached, instabilities in the positive part of the 
spectrum do not appear, even for large values of space charge.
The only instabilities are in the negative part of the spectrum,
vanishing with an increment of SC, as expected.

%------------------------------------------------------------------------------%
In the SSCHP model there is no instability as it was in the case of the sine 
wake;
the mode crossing observed in the situation of the sine wake turns into an
approach-divert behavior and the modes separate when $\alpha$ is increased.
%==============================================================================%
%------------------------------------------------------------------------------%
%------------------------------------------------------------------------------%
%------------------------------------------------------------------------------%
%==============================================================================%
\section{\label{sec:RnD}Summary}

%------------------------------------------------------------------------------%
We questioned behavior of the transverse mode coupling 
thresholds as functions of space charge, 
wakes, potential wells and bunch distribution functions. Two different 
analytical models describing the bunch with constant line density
had been used.
The first one is the airbag model described in the article published by 
M.~Blaskiewicz almost 20 years ago~\cite{blaskiewicz1998fast}.
This model covers all values of space charge and yields an eigenvalue problem 
which is easily solved to machine precision for arbitrary combinations of the 
exponential and sinusoidal wakes.
A complimentary model is given by the SSC 
theory~\cite{burov2009head,burov2015damping,balbekov2009transverse}.
In contrast to the ABS, it can be used only at sufficiently high SC tune shift, 
but it can
be applied for any shape of the potential well and the bunch distribution 
function.
As examples of the SSC models, we took an arbitrary bunch in a square well and 
the
Hofmann-Pedersen distribution $f_\text{hp}^{(0)}$ for the parabolic potential 
well.
They both yield analytical solutions, allowing to compute a large number of the 
wake matrix elements with extremely high accuracy, thus helping to distinguish 
pure numerical thresholds from the real ones. 
It has been demonstrated that in the SSC limit, Eq.~(\ref{math:SSC}), all the 
models give similar results, while the SSCSW fully coincides with the ABS, as 
expected.
The SSCHP model demonstrates good quantitative agreement with the other two, 
except for a
special case of a sine wake, where, for this distribution function, the modes 
may cross 
but never couple.
In Section~\ref{sec:NegWakes} we investigated negative wakes; our results agree 
with similar ones of M.~Blaskiewicz and V.~Balbekov: the TMCI thresholds tend 
to 
go linearly with the SC parameter, thus vanishing in the SSC case. 
%The constant and exponential wakes were considered first;
%we summarized results and made a comparison between the models.
%In addition, the thick resistive wall wake was considered for the SSCSW and 
the 
%SSCHP cases.
%For all negative wakes the instability happens only when the wake-driven and 
SC 
%tune shifts are comparable, and, occurs only at the negative part of the 
%spectrum;
%Space charge helps against the TMCI by deflecting $\Delta Q_{-1}$ mode down 
and 
%thus preventing its coupling with zeroth mode, deflected down by a wake, 
%\cite{ng1999stability}.
%The instability threshold monotonically increases with the space charge and the
%TMCI
%vanishes at the SSC limit~\cite{blaskiewicz1998fast}.
%This result was confirmed for both SSCSW and SSCHP models: the threshold 
%monotonically increases with the mode truncation parameter showing a beam 
%stability.
%
%------------------------------------------------------------------------------%

The situation is richer with 
possibilities for the oscillating wakes.
First, an increase of TMCI threshold is non-monotonic here and, for sufficiently 
high SC 
parameter, TMCI threshold goes down inversely proportional to the SC tune shift, 
according to the hypothesis of Ref.~\cite{burov2009head}:
in addition to vanishing TMCIs in the negative part of the spectrum, there are 
non-vanishing TMCIs associated with couplings of the positive modes. 

%------------------------------------------------------------------------------%
For the sine wake, three special features were seen.
The first one is that when SC is zero, there is no instability for positive 
modes independent of the wake phase advance;
but, when the SC is increased, a cascade of positive mode couplings and 
decouplings 
appear.
These multiple couplings-decouplings for larger values of wake amplitude is the 
second 
major
distinction.
The last difference is that the HP model has a unique feature ---
there is no instability in SSC limit for the sine wake.
It looks like an unknown theorem prohibits coupling for any value of wake 
amplitude for this case --- 
one only observes the mode crossing or approach-divert behavior instead.

%------------------------------------------------------------------------------%
Based on our numerical results, we will conclude that the behavior of the TMCI 
thresholds with the SC parameter depends on the nature of wake function, 
roughly, on the number of its oscillations per bunch.
When the wake function is negative, the TMCI is of the vanishing type: its 
threshold grows linearly with the SC parameter when the latter is high enough;
the same is true for "effectively'' negative wakes, i.e. oscillating wakes with 
moderate phase advance,
$\omega\,\tb < \pi$, or the broadband wake with a sufficiently strong 
decay rate.
For the oscillating wakes, the threshold dependence on the SC parameter should 
be expected as non-monotonic, going inversely proportional to the SC parameter 
when it is high enough.

%------------------------------------------------------------------------------%
% Based on these results some practical recommendations can be suggested for
% machines with dominating space charge.
% If the wake function has about one oscillation per bunch, then reduction of a 
% bunch length (alternatively suppression of oscillation of a wake) should 
% eliminate TMCI when wake can be considered as a negative.
% If wake have many oscillations per bunch and significant reduction of length 
% is not possible, extension of a bunch length (or alternatively modification of
% wake function in order to have more oscillations per bunch) should increase 
% the TMCI threshold.

%==============================================================================%
%==============================================================================%
%==============================================================================%

\begin{acknowledgments}

%------------------------------------------------------------------------------%
This manuscript has been authored by Fermi Research Alliance, LLC under 
Contract 
No. DE-AC02-07CH11359 with the U.S. Department of Energy, Office of Science, 
Office of High Energy Physics. The U.S. Government retains and the publisher, 
by 
accepting the article for publication, acknowledges that the U.S. Government 
retains a non-exclusive, paid-up, irrevocable, world-wide license to publish or 
reproduce the published form of this manuscript, or allow others to do so, for 
U.S. Government purposes.
%The author would like to thank
%{\bf Leo~Michelotti},
%{\bf Eric~Stern}
%and
%{\bf James~F.~Amundson}
%for their discussions and valuable input.
%{\bf Alexey~Burov} for encouraging to find full family of solutions.
%{\bf Valeri~Lebedev} whose solution for electrostatic quadrupole led me to
%generalization, just as in the case with {\bf E.~M.~McMillan} and
%{\bf F.~Krienen}.
%And, of course, {\bf Sergei~Nagaitsev} who brought back to life original
%unknown McMillan's article which helped me with symmetric description of
%electromagnetic fields.
\end{acknowledgments}

%==============================================================================%
%==============================================================================%
%==============================================================================%
%==============================================================================%
%==============================================================================%
\appendix

%==============================================================================%
%------------------------------------------------------------------------------%
%------------------------------------------------------------------------------%
%------------------------------------------------------------------------------%
%==============================================================================%
\section{\label{secAP:Wlm}Matrix elements $\mW$ for SSC models}

Wake function $W(\tau)$ and corresponding matrix elements $\mW$ for the SSCHP 
model are provided below in a Table~\ref{tab:WlmHP}.
Calculation of matrix elements for SSCSW model boils down to standard integrals 
of two trigonometric and an exponential functions.
They are easy to compute and the reader can find them using almost any symbolic 
computation program;
matrix elements for sine and cosine wake functions should be calculated with 
care when $\omega\,\tb = \pi\,n$ with $n=2,3,4,\ldots$ due to a resonance 
condition.

%==============================================================================%
\begin{table*}[p!]
\caption{\label{tab:WlmHP}
	Matrix elements $\mW$ for SSCHP model with negative (delta function,
	constant, exponential and resistive wall) and oscillating wakes 
	(trigonometric and broadband impedance).
	$\mW$ for all oscillating wakes are expressed using matrix elements of 
	exponential wake.
	}
\begin{ruledtabular}
\begin{tabular}{ll}
$W(\tau),\,\tau \leq 0$				& $\mW$		\\[-0.25cm]
								\\\hline
								\\[-0.1cm]
\multicolumn{2}{c}{Negative wakes}				\\[0.25cm]
$-W_0\delta(\tau)$				&
$\displaystyle-\frac{W_0}{2}\,\delta_{l,m}$			\\[0.35cm]
$-W_0$						&
$\displaystyle-\frac{W_0}{2}
	\left[
		\delta_{lm,l-m} -
	(-1)^{\left\lfloor{l/2}\right\rfloor + \left\lfloor{m/2}\right\rfloor}
		\frac{\delta_{l,m+1} - \delta_{l,m-1}}
		{\sqrt{(2\,l+1)(2\,m+1)}}
	\right]	$						\\[0.35cm]
$-W_0\exp(\alpha\,\tau)$			&
$\displaystyle-W_0\sum_{lm}^{ij}
\frac{j!}{\alpha^{j+1}}
\sum_{k=0}^j \frac{\alpha^k}{k!}
\left\{
\frac{1+(-1)^{i+k}}{i+k+1} - \frac{e^{-\alpha}}{(-\alpha)^{i+1}}
\left[ \Gamma(i+1,\alpha) - \Gamma(i+1,-\alpha) \right]
\right\}$
\footnote{Where operator $
\sum_{lm}^{ij} =
(-1)^{\left\lfloor{l/2}\right\rfloor+\left\lfloor{m/2}\right\rfloor}\,
2^{l+m-2}\,\sqrt{(2\,l+1)(2\,m+1)}\,\sum_{i=0}^l\,\sum_{j=0}^m
	\begin{bmatrix}
		l		\\ i
	\end{bmatrix}
	\begin{bmatrix}
		m		\\ j
	\end{bmatrix}
	\begin{bmatrix}
		\frac{l+i-1}{2}	\\ l
	\end{bmatrix}
	\begin{bmatrix}
		\frac{m+j-1}{2}	\\ m
	\end{bmatrix}
$.}								\\[0.35cm]
$-W_0/\sqrt{|\tau|}$				&
$\displaystyle-W_0
	\sum_{lm}^{ij}\frac{2^{i+j+\frac{3}{2}}}{i+j+\frac{3}{2}}\,
	\sum_{k=0}^j \frac{(-1)^{i+k}}{j-k+\frac{1}{2}}\,
	\begin{bmatrix}
	j \\ k
	\end{bmatrix}\,
	_{2}F_{1}\left[
		-\left( i+j+\frac{3}{2} \right);
		-\left( i+k             \right);
		-\left( i+j+\frac{1}{2} \right);
		\frac{1}{2}
	\right]$						\\[0.65cm]
\multicolumn{2}{c}{Oscillating wakes}				\\[0.25cm]
$-W_0\cos(\omega\,\tau)$			&
$ \Re\left[\mW^\mathrm{exp}(i\,\omega)\right]$			\\[0.45cm]
$ W_0\sin(\omega\,\tau)$			&
$-\Im\left[\mW^\mathrm{exp}(i\,\omega)\right]$			\\[0.45cm]
$ W_0\sin(\omega\,\tau)\,\exp(\alpha\,\tau)$	&
$-\Im\left[\mW^\mathrm{exp}(\alpha+i\,\omega)\right]$
\end{tabular}
\end{ruledtabular}
\end{table*}
%==============================================================================%

%==============================================================================%
%------------------------------------------------------------------------------%
%------------------------------------------------------------------------------%
%------------------------------------------------------------------------------%
%==============================================================================%
\section{\label{secAP:AB}Exponential and trigonometric wakes for ABS model}

%------------------------------------------------------------------------------%
Matrix $\mathrm{M}$ for exponential, cosine and broadband impedance wakes is 
provided below in a Table~\ref{tab:MAB}.

%==============================================================================%
\begin{table*}[h]
\caption{\label{tab:MAB}
	Matrix $\mathrm{M}$ for ABS model with exponential (or constant for 
	 $\alpha = 0$), cosine and broadband impedance (or sine for $\alpha=0$) 
	wake functions.
	}
\begin{ruledtabular}
\begin{tabular}{lc}
$W(\tau),\,\tau\leq 0$	& $\mathrm{M}$	\\[-0.25cm]
				\\\hline
				\\[-0.1cm]
$-W_0\,e^{\alpha\,\tau}$				&
$
\begin{bmatrix}
 i\,\pi\left( \frac{1}{2}\frac{\Qsc}{\Qs} + \frac{\Qx}{\Qs} \right)	&
-i\,\pi\,\frac{1}{2}\frac{\Qsc}{\Qs}					&
 i\,\pi									\\
 i\,\pi\,\frac{1}{2}\frac{\Qsc}{\Qs}					&
-i\,\pi\left( \frac{1}{2}\frac{\Qsc}{\Qs} + \frac{\Qx}{\Qs} \right)	&
-i\,\pi									\\
-\frac{1}{2}\frac{\kappa\,W_0\tb}{\Qs}				&
-\frac{1}{2}\frac{\kappa\,W_0\tb}{\Qs}				&
 \alpha\,\tb								\\
\end{bmatrix}
$							\\[1.cm]
$-W_0\,\cos(\omega\,\tau)$				&
$
\begin{bmatrix}
 i\,\pi\left( \frac{1}{2}\frac{\Qsc}{\Qs} + \frac{\Qx}{\Qs} \right)	&
-i\,\pi\,\frac{1}{2}\frac{\Qsc}{\Qs}					&
 i\,\pi									&
 i\,\pi									\\
 i\,\pi\,\frac{1}{2}\frac{\Qsc}{\Qs}					&
-i\,\pi\left( \frac{1}{2}\frac{\Qsc}{\Qs} + \frac{\Qx}{\Qs} \right)	&
-i\,\pi									&
-i\,\pi									\\
-\frac{1}{4}\frac{\kappa\,W_0\tb}{\Qs}				&
-\frac{1}{4}\frac{\kappa\,W_0\tb}{\Qs}				&
 i\,\omega\,\tb								&
 0									\\
-\frac{1}{4}\frac{\kappa\,W_0\tb}{\Qs}				&
-\frac{1}{4}\frac{\kappa\,W_0\tb}{\Qs}				&
 0									&
-i\,\omega\,\tb
\end{bmatrix}
$							\\[1.cm]
$ W_0\,\sin(\omega\,\tau)\,e^{\alpha\,\tau}$		&
$
\begin{bmatrix}
 i\,\pi\left( \frac{1}{2}\frac{\Qsc}{\Qs} + \frac{\Qx}{\Qs} \right)	&
-i\,\pi\,\frac{1}{2}\frac{\Qsc}{\Qs}					&
 i\,\pi									&
 i\,\pi									\\
 i\,\pi\,\frac{1}{2}\frac{\Qsc}{\Qs}					&
-i\,\pi\left( \frac{1}{2}\frac{\Qsc}{\Qs} + \frac{\Qx}{\Qs} \right)	&
-i\,\pi									&
-i\,\pi									\\
-\frac{i}{4}\frac{\kappa\,W_0\tb}{\Qs}				&
-\frac{i}{4}\frac{\kappa\,W_0\tb}{\Qs}				&
 (\alpha+i\,\omega)\,\tb							
	&
 0									\\
 \frac{i}{4}\frac{\kappa\,W_0\tb}{\Qs}				&
 \frac{i}{4}\frac{\kappa\,W_0\tb}{\Qs}				&
 0									&
 (\alpha-i\,\omega)\,\tb
\end{bmatrix}
$
\end{tabular}
\end{ruledtabular}
\end{table*}
%==============================================================================%

%\nocite{*}

%\bibliography{bibfile}			% Produces the bibliography via BibTeX.

%

\end{document}